\documentclass[aps,prb,twocolumn,floatfix,footinbib]{revtex4-1}

\usepackage{epsfig}
\usepackage{bm}
\usepackage{amsfonts}
\usepackage{amsmath}
\usepackage{amssymb}
\usepackage{graphicx}
\usepackage{color}
\usepackage{mathrsfs} 
\usepackage[normalem]{ulem}
\begin{document}

\preprint{\today}

\title{Nodal ``ground states'' and orbital textures in semiconductor quantum
dots}

\author{Jeongsu Lee$^{1}$}
\author{Karel V\'{y}born\'{y}$^{1,2}$}
\author{Jong E. Han$^{1}$}
\author{Igor \v{Z}uti\'{c}$^{1}$}

\affiliation{
$^{1}$ Department of Physics, State University of New York 
at Buffalo, NY, 14260, USA \\
$^{2}$ Institute of Physics ASCR, v.v.i., Cukrovarnick\'{a} 10, CZ-16253, 
Praha 6, Czech Republic}

\date{\today}

\begin{abstract}
Conventional understanding implies that the ground state of a nonmagnetic
quantum mechanical system should be nodeless. While this notion also provides
a valuable guidance in understanding the ordering of energy levels in
semiconductor nanostructures, there are reports that {\em nodal} ground states
for holes are possible.  However, the existence of such  nodal states has been
debated and even viewed merely as an artifact of a $\bm{k{\cdot}p}$ model.
Using complementary approaches of both $\bm{k{\cdot}p}$ and tight-binding
models, further supported by an effective Hamiltonian for a continuum model,
we reveal that nodal ground states in quantum dots are not limited
to a specific approach. Remarkably, the emergence of nodal hole states at the
top of the valence band can be attributed to the formation of orbital vortex
textures through competition between the hole kinetic energy and the coupling 
to the conduction band states.
We suggest an experimental test for our predictions of the reversed energy
ordering and the existence of nodal ground states. We discuss how our findings and
the studies of orbital textures could be also relevant for other materials systems.
\end{abstract}

\maketitle

\section{Introduction}

Unlike the common expectation that a bound state of a particle should be 
nodeless,\cite{Marder:2010,Snoke:2008,Gywat:2010} 
theoretical calculations in semiconductor quantum wires\cite{Persson2006:PRB} and quantum 
dots\cite{Bagga2003:PRB,Bagga2005:PRB,Bagga2006:PRB,Yu2003:JPC} (QDs) have predicted 
hole ground states with a node. Those ground states occur with the inversion of the 
energy level ordering between nodeless ($S$-like) and nodal ($P$-like) wavefunctions 
due to various factors, such as the confinement size and strength, the choice of a material, 
and the spin-orbit interaction. 
This peculiar phenomenon is connected to the 
formation of  dark excitons that exhibit very long recombination time and
Stokes shift\cite{Efros1995:PRL,Efros1996:PRL,Bagga2006:PRB} in the luminescence of 
InAs and CdSe QDs. 
One can expect that such nodal states would also 
have intriguing implications for magnetically-doped QDs and provide
additional control for their magnetic
ordering.\cite{Seufert2001:PRL,Besombes2004:PRL, Leger2005:PRL,Henneberger:2010, 
Fernandez-Rossier:2011, Beaulac2009:S,%
 Xiu2010:ACSN,Bussian2009:NM,Sellers2010:PRB, Fernandez2004:PRL, Zhang2007:PRB, 
 Govorov2005:PRB,Abolfath2008:PRL, Oszwaldowski2011:PRL,Pientka2012:PRB}
In this work we show how the presence of the nodal ground
state is associated with the emergence of orbital textures that minimize the
energy of underlying model we use to describe QDs. 

While there appeared some criticism asserting that the existence of nodal ground state is just a 
theoretical artifact,\cite{Fu1997:APL,Wang1998:JPCB,Wang2000:APL} a systematic effort 
to either prove or disprove the occurrence of the level ordering inversion between the
nodal and nodeless states is still missing. In fact, one is tempted to invoke different
arguments to dismiss the occurrence of such nodal ground states. Elementary
understanding of quantum mechanics would suggest that the nodeless state and 
thus minimization of the kinetic energy should be preferred. A similar reasoning 
would follow from the Sturm-Liouville theorem for 
differential equations.\cite{Morse:1953} 

\begin{figure}[htbp]
\includegraphics[scale=0.35]{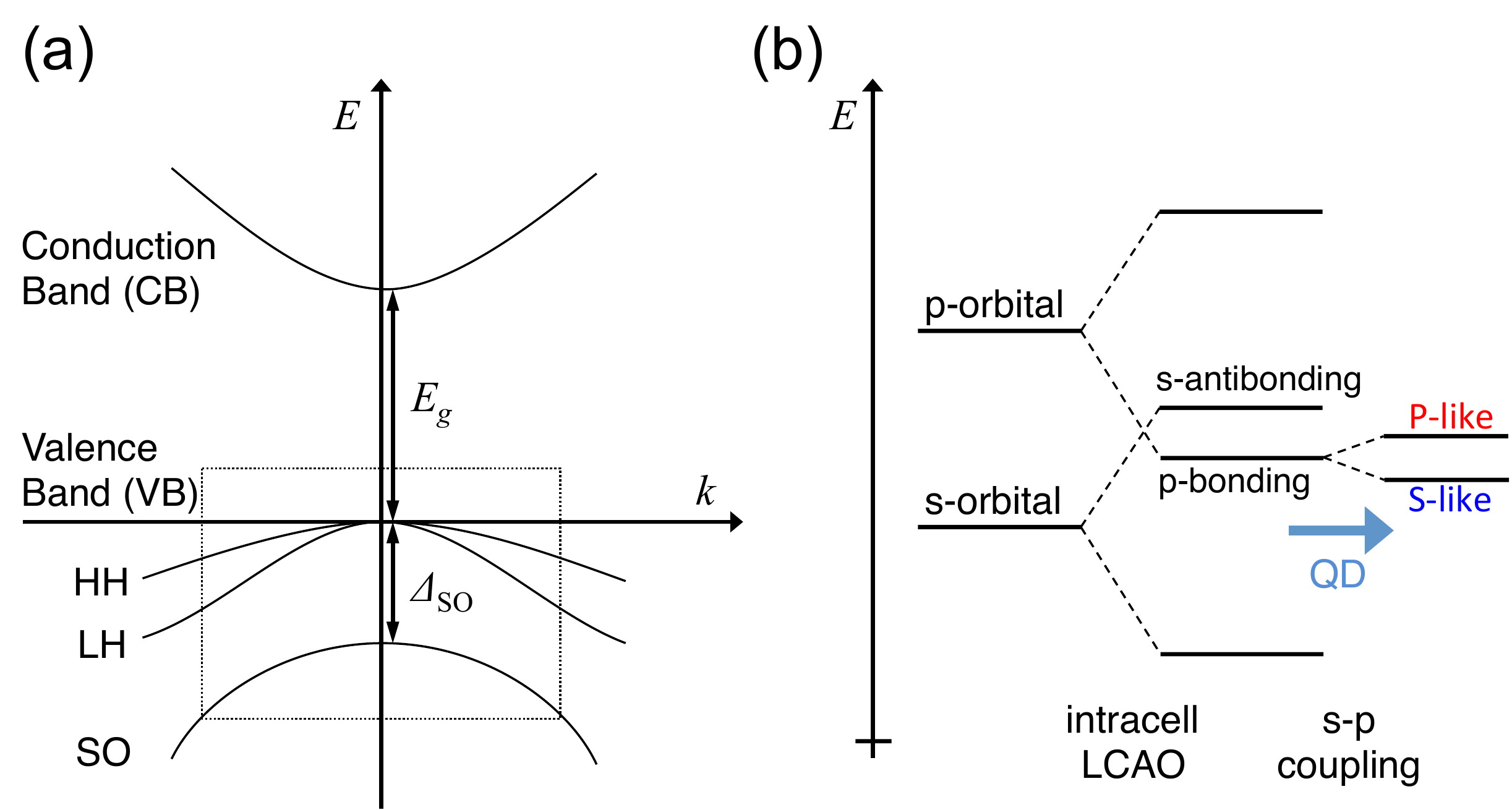}
\caption{A qualitative sketch of a diamond or zinc-blende lattice 
semiconductor bulk band structure in the 
$\bm{k{\cdot}p}$  and  tight-binding model.
(a) Conduction and valence bands are separated by the energy gap, $E_g$.
The valence band comprises heavy and light hole (HH,LH) bands,  
separated at the $\Gamma$ point ($\bm{k}=0$) by the spin-orbit gap,
$\Delta_{\rm{SO}}$, from the split-off (SO) band. 
The six-band $\bm{k{\cdot}p}$ Hamiltonian describes the region 
marked by the box. (b) Within the two atom basis of Bravais lattice, 
linear combinations of atomic orbitals (LCAOs) form the $s$-antibonding and $p$-bonding orbitals. The latter give rise to valence band once the gap in the bulk band structure opens due to $s$-$p$ hybridization. This $s$-$p$ coupling also influences the ordering of QD levels. If it is strong enough, the $P$-like (nodal) envelope function state appears at the top of the valence band.}\label{fig:01}
\end{figure}

However, a closer look at the character of the hole ground state 
displays more complexities, which prevent
us from simply concluding that
a nodal wavefunction cannot describe the ground state.
The hole ground state is not a ``true" 
ground  state, but a state in the middle of the whole spectrum
of single-particle energy levels.
The hole represents a lack of an electron, the ground state of hole 
refers to the highest energy state in the valence band (VB) in the electronic energy dispersion.
Additionally, the corresponding Hamiltonian describes multiband wavefunctions 
arising from the band structure as shown in Fig.~\ref{fig:01}(a).
Therefore, the hole ground state with a nodal wavefunction is not forbidden. 

Even if the suggested nodal ground states are not impossible, are the
methods employed to identify them indeed appropriate? 
Specifically, their identification
relies on the use of a $\bm{k{\cdot}p}$ model 
which is commonly employed to describe
the band structure in bulk semiconductors. 
The model is based on an effective Hamiltonian with 
periodic Bloch functions as its 
solutions which, for a certain crystal momentum $\bm{k}$, form a complete set. 
Near a specific $\bm{k}$ (typically $\Gamma$-point,  $\bm{k}=0$)
an accurate band structure is obtained perturbatively.\cite{Yu:2010}

Several qualitative features arising from a  $\bm{k{\cdot}p}$
model applied to diamond or zinc-blende semiconductors, such as the opening of the 
energy gap, $E_g$, and the VB structure are illustrated in Fig.~\ref{fig:01}(a). 
In the VB, the heavy and light holes 
are separated by the spin-orbit gap, $\Delta_{\rm{SO}}$, from the split-off band.
The versatility of the  $\bm{k{\cdot}p}$ model 
was successfully used to elucidate a wealth of phenomena in semiconductors and their
nanostructures, including optical properties,\cite{Winkler:2003,Gywat:2010,
Meier:1984,Dyakonov:2008,Drouhin2008:PRB} the spin 
Hall effect,~\cite{Murakami2003:S} topological insulators,\cite{Hasan2010:RMP}
and \emph{Zitterbewegung}.\cite{Bernardes2007:PRL}
While a $\bm{k{\cdot}p}$ model   with an envelope function approximation is also 
frequently applied to low-dimensional systems, such as quantum wells, wires, and 
dots,\cite{kdotp,Winkler:2003,Gywat:2010}
sometimes it is more challenging  to justify its validity. 

A complementary picture to examine the occurrence of  the nodal 
ground  states can be obtained from a tight-binding (TB) model. 
Unlike the $\bm{k{\cdot}p}$ model, one can consider an electron in a solid 
as localized within an isolated atom placed at each lattice site. If no 
interaction between neighboring atoms exists, there are degenerate energy 
levels corresponding to $s$- and $p$-orbitals, shown on the left in Fig.~1(b). 
In a realistic solid, since atoms are not completely isolated, an electron at 
one site interacts with the neighboring atoms. In a TB model, electronic 
wavefunction in the primitive unit cell can be approximated by a linear 
combination of atomic orbitals (LCAO) of isolated atoms at different sites. 
If we next consider hopping among $s$- and $p$-orbitals, bonding and 
antibonding states will form, Fig.~1(b), middle. Finally, when we introduce 
$s$-$p$ coupling, the semiconducting gap opens in the bulk.  Such an 
atomistic TB model provides a different approach to study semiconductor 
nanostructures in which the $\bm{k{\cdot}p}$ model may not capture the 
full symmetry of atomic wavefunctions, and it could help our microscopic 
understanding of the level ordering in a QD.

We combine these two complementary models to systematically explore the presence 
of nodal  ground states and to eliminate the possibility that our results are an artifact of a 
specific method.  Within both  $\bm{k{\cdot}p}$ and TB models  we predict an 
unconventional level ordering in which the nodal ($P$-like) wavefunction can attain an 
energy lower than the nodeless ($S$-like) wavefunciton. This departure from the conventional 
nodeless ground state can be understood even from a simple TB description and we can attribute it to mixing between orbitals of different types centered at different atomic sites.
We also develop an effective 
Hamiltonian from the continuum limit of a TB model which explains how the emergence of nodal 
ground state is related to the orbital ordering and the formation of orbital textures. 
Throughout this paper, uppercase (lowercase) letters $S$,$P$ ($s$,$p$) refer to envelope functions 
(atomic orbitals).

Our presentation is organized as follows. After this Introduction, in Sec.~II we provide some 
background about the $\bm{k{\cdot}p}$  model and its parameters. We then describe 
the phase diagram for the occurrence of a nodal ground state. In Sec.~III we describe several 
TB models in both two and three dimensional crystal structures and 
identify the microscopic origin of the nodal ground state. In Sec.~IV we discuss  several implications of our findings and 
propose an experiment which could be used to probe the presence of nodal ground states in
QDs. Our conclusions provide the main findings of this work as well as the 
possible future directions.

\section{$\bm{k{\cdot}p}$  Model}

Formulation of a bulk $\bm{k{\cdot}p}$ model can vary significantly
in its complexity, the choice of the specific  system, and the number of 
bands included. For transparency, we focus on its main features in a simple
implementation of the non-interacting single electron
picture. Invoking the Bloch theorem, $\Phi_{n{\bm k}}=
u_{n{\bm k}}(\bm r) e^{i {\bm k\cdot r}}$,
we obtain an effective Schr\"{o}dinger equation for Bloch functions 
\begin{equation}
(\hat{H}_0+\hat{H}_1)u_{n{\bm k}}=
E_{n{\bm k}}u_{n{\bm k}},
\label{eq:kpS}
\end{equation}
with $\hat{H}_0=\hat{p}^2/2m+V$ and $\hat{H}_1=\hbar
\bm{\hat{k} \cdot \hat{p}}/m+\hbar^2k^2/2m$, where $n$ is the
the band index, $\hat{p}$ momentum operator, the Bloch function
$u_{n{\bm k}}$ has the periodicity of the lattice potential
$V$, and $m$ is the free electron mass. If $u_{n{\bm 0}}$ and $E_{n{\bm 0}}$ 
are known at $\Gamma$ point, $\hat{H}_1$ can be 
treated as a perturbation in terms of ${\bm k}$
suggesting that for QDs, the $\bm{k{\cdot}p}$ model should work 
better for larger systems where $k$ is smaller. To some extent, this
limitation can be removed by using a large number of bands so that the
bulk $\bm{k{\cdot}p}$ model accurately reproduces the band
structure over the whole Brillouin zone. However, there is only a 
limited number of experimentally determined energy gaps and
matrix elements which are needed as the input to the model.\cite{Yu:2010}

We begin by
considering a reduced Hilbert space that consists of heavy hole (HH) and light
hole (LH) bands. The $\bm{k{\cdot}p}$ perturbative procedure
leads to the Luttinger Hamiltonian,\cite{Luttinger,Winkler:2003,Yu:2010} given 
in terms of  the Kohn-Luttinger parameters $\gamma_1$, $\gamma_2$, 
and $\gamma_3$. The standard procedure for confined
systems,\cite{Winkler:2003} i.e., the replacement 
$\bm{k}\mapsto-i\nabla$, leads to
\begin{equation}
\begin{split}
H_L=&\frac{\hbar^2}{2m} \bigg[\bigg(\gamma_1+\frac52\gamma_2\bigg)\nabla^2
-2\gamma_3({\bm \nabla} \cdot {\bm \hat{J}})^2 \\
&+2(\gamma_3-\gamma_2)(\nabla_x^2\hat{J}_x^2+c.p.)\bigg],
\label{eq:lutt}
\end{split}
\end{equation}
where ${\bm \hat{J}}$ is the $J=3/2$ angular momentum operator, and
$c.p.$ stands for cyclic permutations. Each set of
$\gamma_1$, $\gamma_2$, $\gamma_3$ represents a specific material, 
and several examples are given in Table~\ref{tab:01}. The four-band model
of Eq.~(\ref{eq:lutt}) corresponds to $\Delta_{\rm{SO}}\to\infty$ limit and we
often employ the spherical approximation in which both $\gamma_2$ and
$\gamma_3$ are replaced by\cite{Gywat:2010,Baldereschi1973:PRB}
\begin{equation}
\tilde{\gamma}_2=(2\gamma_2+3\gamma_3)/5.
\label{eq:tilde}
\end{equation}
This approximation suppresses the warping (anisotropy) of Fermi surface in the bulk
and the effective masses are then $m_{\rm{HH}}=m/(\gamma_1 -
2\tilde{\gamma}_2)$ and $m_{\rm{LH}}=m/(\gamma_1 +
2\tilde{\gamma}_2)$.\cite{Gywat:2010}
VB energy levels for QDs can
be calculated by diagonalization of Eq.~(\ref{eq:lutt}) expressed in a
suitably chosen basis in $\bm{k}$-space for the envelope functions.\cite{note2}
We consider a cubic domain of dimensions $L_x=L_y=L_z\equiv L$
with zero boundary conditions unless explicitly mentioned otherwise.
The cubic geometry simplifies the comparison between $\bm{k{\cdot}p}$ and TB models,
however, choosing a different confinement geometry, e.g., spherical or cuboid 
(see Appendix D) confinement, does not  
change qualitatively our findings, as long as the geometry is highly symmetric.

\begin{table}
\caption{Kohn-Luttinger parameters  
$\gamma_1$, $\gamma_2$, $\gamma_3$
 and the spin-orbit gap $\Delta_{\rm{SO}}$ of selected 
materials.\cite{Winkler:2003}  For the spherical approximation to
Eq.~(\ref{eq:lutt}), it holds $\tilde{\gamma}_2=(2\gamma_2+3\gamma_3)/5$.}
\label{tab:01}
\begin{tabular}{c|cccc|c}
\hline\hline
Material & $\gamma_1$ & $\gamma_2$ & $\gamma_3$ &  
  $\Delta_{\rm{SO}}$(meV)  & $\tilde{\gamma}_2/\gamma_1$ \\
\hline
GaAs & 6.85 & 2.10 & 2.90 & 341 & 0.377 \\
InAs & 20.40 & 8.30 & 9.10 & 380 & 0.430 \\
CdTe & 5.30 & 1.70 & 2.00 & 949 & 0.355 \\
Ge & 13.38 & 4.24 & 5.69 & 290 &  0.382\\
\hline
\end{tabular}
\end{table}

\begin{figure}[tbp]
\includegraphics[scale=0.5]{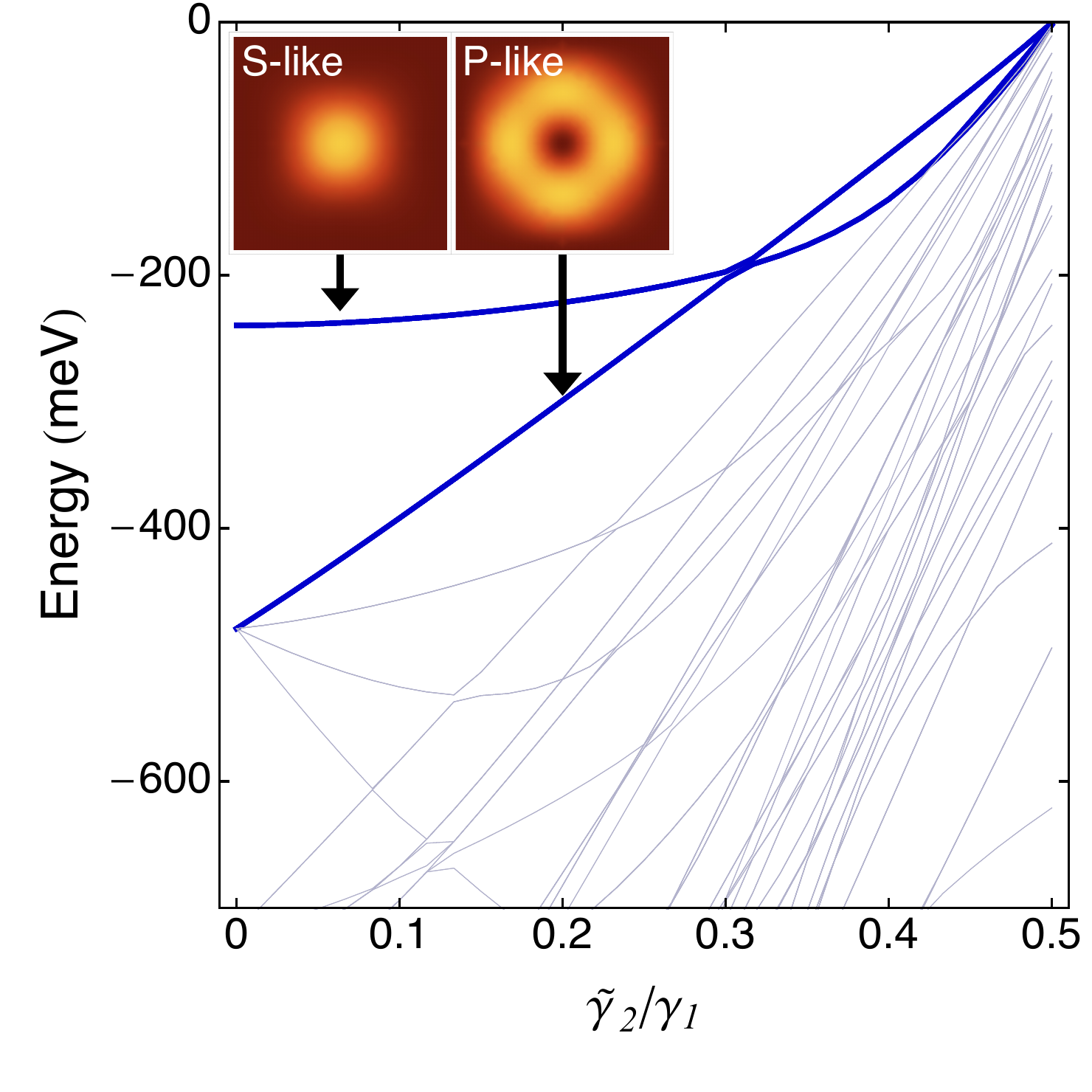}
\caption{Hole energy spectrum of a cubic QD for $\Delta_{\rm{SO}}=0$ 
under the spherical approximation. 
The chosen characteristic energy scale for the Luttinger Hamiltonian 
$\varepsilon_0=16.2$ meV
corresponds to an InAs QD 
($L=9.8$~nm)\cite{Kash1992:PRL} when $\tilde{\gamma}_2/\gamma_1=0.430$.
Zero energy is placed at the top of the VB in the bulk.
Different Kohn-Luttinger parameters $\tilde{\gamma}_2/\gamma_1$
can be interpreted as representing different material choices.
Squared moduli $|\Psi(x,y,z)|^2$, $z=L/2$, of the envelope functions of the two
uppermost levels (not counting the degeneracy) are shown as insets.
The crossing of these two sixfold degenerate levels delimits the region 
of the $P$-like ground state: $\tilde{\gamma}_2/\gamma_1>0.31$.}\label{fig:02}
\end{figure}

The four-band model described by Eq.~(\ref{eq:lutt}) can be generalized 
to six bands by including two split-off bands of the VB 
and spin-orbit interaction
\begin{eqnarray}
\hat{H}_{\rm{SO}}=2\frac{\Delta_{\rm{SO}}}{3}{\hat{{\bm S}}\cdot \hat{{\bm L}}},
\end{eqnarray}
where $\hat{{\bm S}}$ and $\hat{{\bm L}}$ are spin and orbital angular momentum 
operators, and $\Delta_{\rm{SO}}$ is the splitting
shown in Fig.~\ref{fig:01}(a).
Instead of the $4\times 4$ matrix in Eq.~(\ref{eq:lutt}), 
we then use standard six-band model as in Eq.~(A8) from
Ref.~\onlinecite{Abolfath2001:PRB} with $\bm{k}\mapsto -i\nabla$
replaced. In the limit $\Delta_{\rm{SO}}=0$,
the spin-orbit coupling effects are eliminated, and the 6~$\times$~6
Luttinger Hamiltonian can be parameterized by the ratio
$\tilde{\gamma}_2/\gamma_1$ and 
$\varepsilon_0=\hbar^2\gamma_1/(m L^2)$
once the spherical approximation is invoked. 
Cubic QD energy levels are plotted in
Fig.~\ref{fig:02} as a function of this parameter which simulates
a continuous variation of the material in terms of $m_{HH}$ and $m_{LH}$.
In the opposite limit of $\Delta_{\rm{SO}}\to \infty$, our Hamiltonian recovers
analogous results, such as Fig.~2 of Ref.~\onlinecite{Vyborny2012:PRB} which is
obtained with four-band model.\cite{note:sign}
We find a crossing between two types of ground states as 
$\tilde{\gamma}_2/\gamma_1$ increases and it occurs well before the
upper limit of $\tilde{\gamma}_2/\gamma_1=0.5$ is reached. Larger
values of $\tilde{\gamma}_2/\gamma_1$ correspond to materials with
small $m_{LH}/m_{HH}$ and the upper limit maps to
$m_{HH}\to\infty$.

Both energy levels shown by bold lines in Fig.~\ref{fig:02} are
sixfold degenerate and their envelope functions modulus squared is
shown in the upper part of the figure. These results imply that, 
in contrast to the common notion, the topmost energy level in the
VB can be the $P$-like state with a node rather than the $S$-like
state without a node. 
The crossing between these two states occurs at approximately
$\tilde{\gamma}_2/\gamma_1=0.31$. Below this value, $S$-like states
are at the top of the VB above the $P$-like states and 
this ordering is reversed as 
$\tilde{\gamma}_2/\gamma_1$ increases. 
Similar behavior is found in spherical QDs.\cite{Yu2003:JPC}
In the following, we discuss materials 
in which the reversed ordering may occur.

\begin{figure}[tbp]
\includegraphics[scale=0.28]{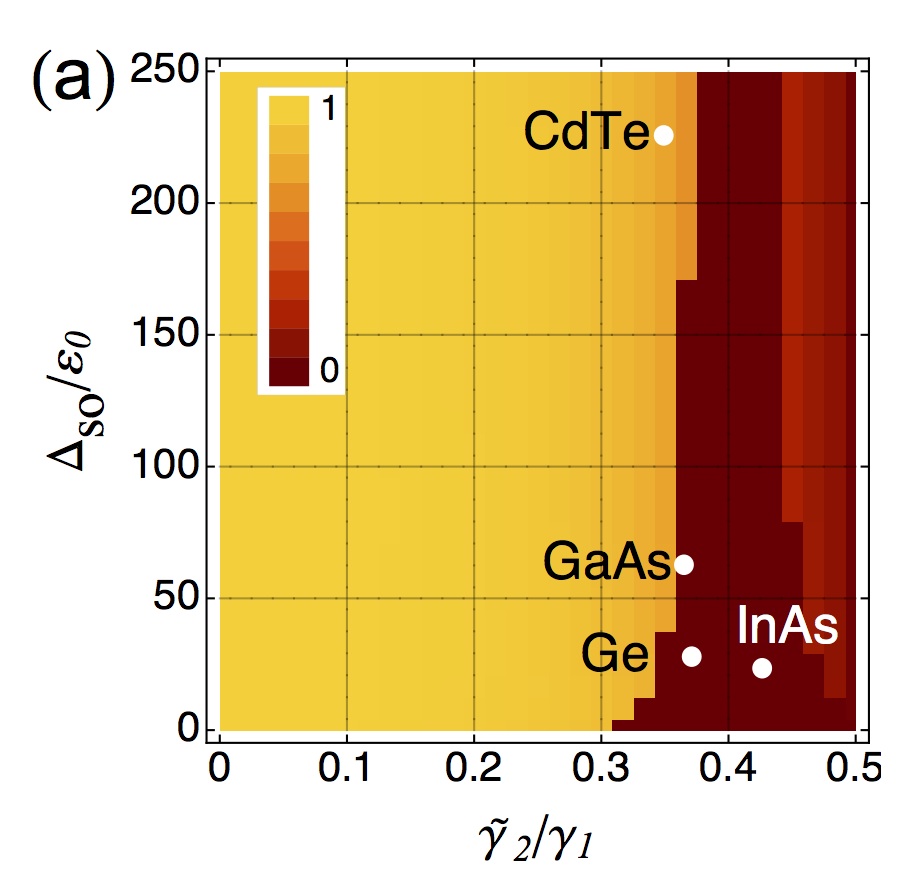}%
\includegraphics[scale=0.28]{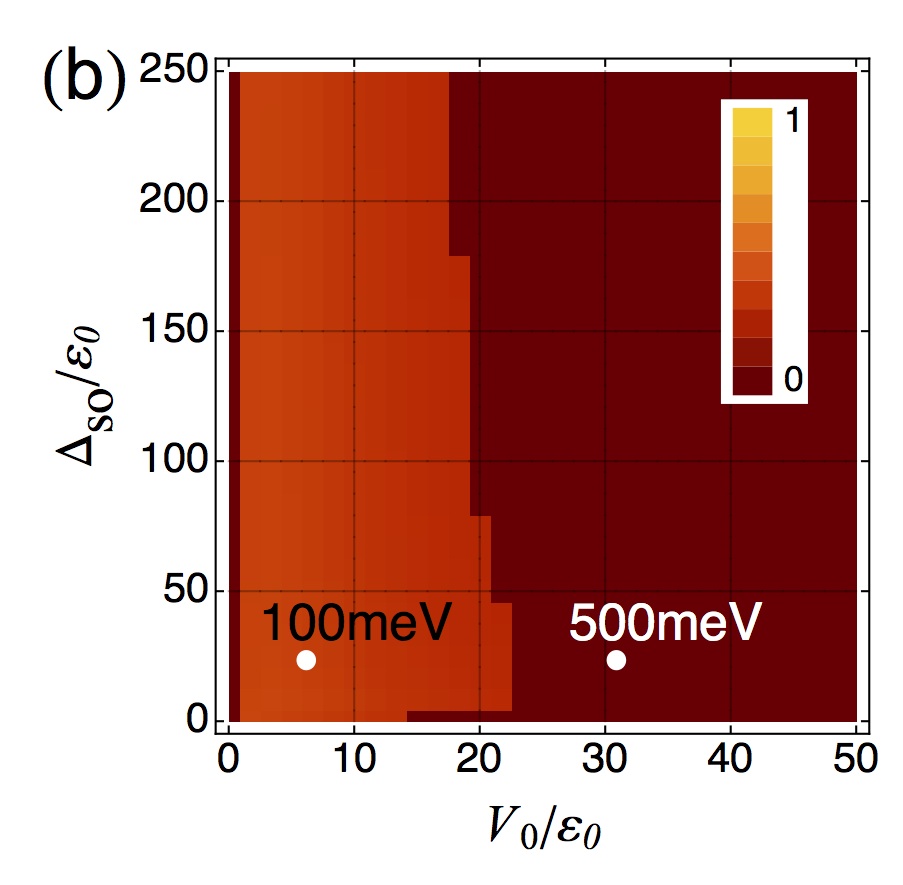}
\caption{QD ground state envelope function projection to the $S$-like states.
Darker areas correspond to $P$-like states.
The ground state character (a) as a function of the Kohn-Luttinger 
parameter ratio $\tilde{\gamma}_2/\gamma_1$ and 
the spin-orbit coupling strength $\Delta_{\rm SO}$;
(b) as a function of the confinement depth $V_0$ and the
spin-orbit coupling strength $\Delta_{\rm SO}$ with InAs QD parameters 
as shown in Table~\ref{tab:01} ($\gamma_2\neq\gamma_3$).
$\Delta_{\rm SO}$ and $V_0$ are normalized with
$\varepsilon_0=\hbar^2\gamma_1/(m L^2)$.
Parameters corresponding to several
typical semiconductors are shown for $L=9.8$ nm.\cite{Kash1992:PRL}
}
\label{fig:03}
\end{figure}

While the ratio $\tilde{\gamma}_2/\gamma_1$ is important, it is not
the only factor that determines the ordering of the two uppermost
states. The crossing of the $S$- and $P$-like states also depends on
other factors, such as the strength of spin-orbit coupling, the depth
$V_0$ and shape of the confinement, and the presence of interfacial
and surface QD states. Even if we employ the spherical approximation
of Eq.~(\ref{eq:tilde}), there are several parameters on which the
energy level ordering depends. We discuss the influence of
$\tilde{\gamma}_2/\gamma_1$, $V_0$ and $\Delta_{\rm{SO}}$ on the topmost VB
level wavefunction character in Fig.~\ref{fig:03}.  The color code
indicates the (squared) weight of the unperturbed $S$-like wavefunction,\cite{note2}
in the darker areas the $P$-like states are closer to the band edge than
$S$-like states.

In Fig.~\ref{fig:03}(a) we assume an infinitely deep confinement,
arguably a good approximation for colloidal QDs.\cite{Gywat:2010} Around
$\tilde{\gamma}_2/\gamma_1=0.4$ there is a window of the reversed
ordering of states, favoring $P$-like ground state.  
Materials such as InAs, GaAs and Ge within this range
are expected to demonstrate the reversed ordering. 
An increase in the normalized spin-orbit coupling,
$\Delta_{\rm{SO}}/\varepsilon_0$ appears detrimental to the
$P$-like ground state. Because of this reason, the well-established
material for single-QD optical experiments, CdTe, misses the range of
reversed level ordering. However, its relatively large value of 
$\Delta_{\rm{SO}}$ could still be compensated for by smaller
dimensions of the QD, since the ordering depends on
$\Delta_{\rm{SO}}/\varepsilon_0\propto L^2$. We also examine the
influence of the finite confinement depth in Fig.~\ref{fig:03}(b) and
consider a well-in-a-well structure. For this panel, we use InAs
parameters given in Table~\ref{tab:01} with $\gamma_2\not=\gamma_3$ and
include a piecewise constant potential $V({\bm r})\propto V_0$ in
our total Hamiltonian.\cite{note2}  
The marked points
are labeled by their confinement depths corresponding to a InAs QD with
$L=9.8$~nm. While deeper confinement makes the reversed
ordering more likely, at a fixed effective depth $V_0/\varepsilon_0$ 
an increase in the spin-orbit coupling can promote $P$-like ground
state, in contrast to what was shown for infinite confinement in 
Fig.~\ref{fig:03}(a).  These
dissimilar trends suggest that the interface between the QD and the
surrounding material or vacuum can alter the ordering of $S$- and
$P$-like states.

We conclude this account of the $\bm{k{\cdot}p}$ model and its
results by recalling that several assumptions,\cite{kdotp,Gywat:2010,Yu:2010}
such as the smoothness
of the confinement potential (on the atomic scale) and sufficiently
small $k$ values involved in the wavefunctions may have not been fully
satisfied in a rigorous manner. It may then be rather puzzling that 
a $\bm{k{\cdot}p}$ model can provide a good agreement with 
experiments,\cite{Gywat:2010} even
when its validity is unclear.

\section{Tight-Binding (TB) Model}

\begin{figure}[htbp]
\begin{centering}
\includegraphics[scale=0.4]{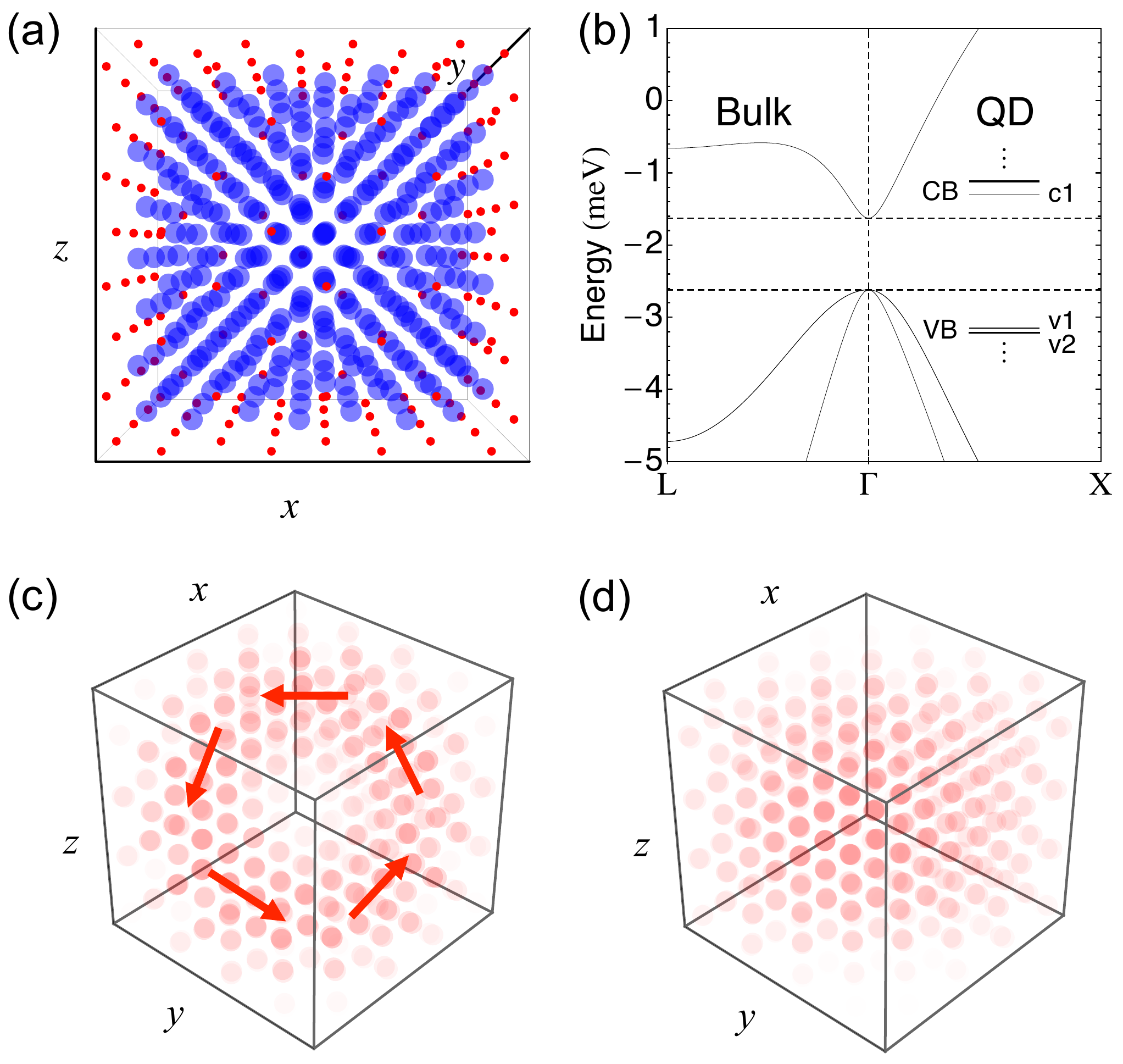}
\par\end{centering}
\caption{TB model results for a Ge QD. 
(a) Diamond lattice structure consisting of 512 atoms in total. 
Large (blue) spheres:  Ge atoms (344); small (red) spheres: passivation layer 
atoms (168). (b) Bulk band structure compared to discrete QD energy levels
(only few levels close to the bulk band gap are shown).
The probability densities for the (c) hole ground state (v1) and the
(d) first excited state of holes (v2), both of them are proportional 
to the color intensity of the spheres. The arrows in (c) describe the 
orbital texture (see the discussion of Fig. 6).}
\label{fig:05}
\end{figure}
To scrutinize the relevance of the $P$-like ground state, 
obtained in Section II using $\bm{k{\cdot}p}$ model, it is important to examine 
if such findings will be preserved within a different framework.
In atomistic TB models, abrupt interface between the QD material and vacuum
is introduced naturally. No special boundary conditions need to be
considered although we still need to pay attention to QD termination and
passivation.
Guided by the phase diagram obtained from the $\bm{k{\cdot}p}$ model 
in Fig.~\ref{fig:03}(a), we now turn to an idealized TB model for Ge QD to 
test the possibility for the $P$-like ground state. Our results, illustrated in 
Fig.~\ref{fig:05}, reveal that such a nodal state indeed appears at the top of the VB and is associated with an orbital texture. In order to understand the origin of the reversed
ordering of energy levels, we discuss
TB models on simpler crystal lattices such as a two-dimensional
(2D) square lattice in Sec.~III A.
We then identify the hybridization between $s$- and $p$-orbitals 
on neighboring atomic sites as the driving mechanism for the
ordering reversal, construct a continuum model which renders this
mechanism clearly understandable and return to the more realistic TB
models to confirm that this mechanism leads to results as shown in
Fig.~\ref{fig:05}.

Our TB models consider only the nearest-neighbor hopping with
one $s$- and three $p$-orbitals at each atomic site. Since
$\bm{k{\cdot}p}$ model in the previous section shows that the
spin-orbit interaction does not promote the occurrence of the $P$-like ground state --- it
even tends to suppress it --- we assume it absent, which would
correspond to $\Delta_{\rm{SO}}=0$ in the notation of Sec.~II. The
spin-up and spin-down states are then degenerate and we can focus
on one of them only.
The TB model is parametrized\cite{Yu:2010} by
the $s$ and $p$-orbital on-site energies,
$E_s=\langle s({\bm R})|\hat{\mathscr{H}}|s({\bm R})\rangle$,
$E_{p}\equiv E_{p_i}=\langle p_i({\bm R})|\hat{\mathscr{H}}|p_i({\bm R})\rangle$, 
where $i=x,y,z$, and four hopping parameters between the nearest-neighbors 
$V_{ss}=\langle s({\bm R}) | \hat{\mathscr{H}} |s({\bm R}+d\hat{{\bm x}}) \rangle$,  
$V_{sp}=\langle s({\bm R}) | \hat{\mathscr{H}} | p_x({\bm R}+d\hat{{\bm x}}) \rangle$, 
$V_{pp\sigma}=\langle p_x({\bm R}) | \hat{\mathscr{H}} | p_x({\bm R}+d\hat{{\bm x}})\rangle$,
and $V_{pp\pi}=\langle p_y({\bm R}) | \hat{\mathscr{H}} | p_y({\bm R}+d\hat{{\bm x}})\rangle$,
where $\hat{\mathscr{H}}$ is the full Hamltonian, ${\bm R}$ is the lattice site, $d$ is the distance between nearest-neighbors.
Note that $\hat{{\bm x}}$ is a unit vector 
along [100] direction, not an operator.
These definitions of hopping parameters are illustrated in Fig.~\ref{fig:06}(a).
In accordance with Slater-Koster rules (see Table~\ref{tab:03}),
the most general TB Hamiltonian we use in this paper can be written as, 
\begin{eqnarray}\label{eq:05}
\hat{H}&=&\sum_{\bm R} E_{\alpha} |\alpha({\bm R})\rangle\langle\alpha({\bm R})|\nonumber \\
&+&\sum_{\bm R,\bm\delta,\alpha,\alpha'} E_{\alpha\alpha'} |\alpha({\bm R})\rangle\langle\alpha({\bm R+{\bm \delta}})|,
\end{eqnarray}
where ${\bm \delta}$ is a vector pointing from ${\bm R}$ to the nearest-neighbors, 
and $|\alpha({\bm R})\rangle$ is $\alpha\in\{s,p_x,p_y,p_z\}$-orbital state at site ${\bm R}$.
$E_\alpha$ is the on-site energy of $\alpha$-orbital, while 
$E_{\alpha\alpha'}$ are ${\bm \delta}$ dependent energy integrals between 
$\alpha$- and $\alpha'$-orbitals separated by ${\bm \delta}$.

\begin{table}
\caption{The Slater-Koster table of interatomic matrix elements.\cite{Slater1954:PR, Harrison:1989}
$l,m,n$ are direction cosines between nearest neighbors, i.e.,
$(l,m,n)={\bm \delta}/|{\bm \delta}|$, where ${\bm \delta}$ represents 
a vector pointing from the left orbital to the right orbital in the subscript. 
Other matrix elements are found by permutation.
$V_{ss}, V_{pp\sigma}, V_{pp\pi}$, 
and $V_{sp}$ are defined in the text.}
\label{tab:03}
\begin{tabular}{ccl}
\hline\hline
\quad $E_{ss}$ & = & $V_{ss}$ \quad \\
\quad $E_{sx}$ & = & $l V_{sp}$ \quad \\
\quad $E_{xs}$ & = & $-l V_{sp}$ \quad \\
\quad $E_{xx}$ & = & $l^2 V_{pp\sigma}+(1-l^2)V_{pp\pi}$ \quad \\
\quad $E_{xy}$ & = & $ln^2 V_{pp\sigma}-lm V_{pp\pi}$ \quad \\
\quad $E_{xz}$ & = & $ln^2 V_{pp\sigma}-ln V_{pp\pi}$ \quad \\
\hline
\end{tabular}
\end{table}

Based on our $\bm{k{\cdot}p}$ calculations
in Fig.~\ref{fig:03}, we chose a diamond-structure germanium QD to
search for a $P$-like state at the top of the VB. 
Having found parameters of the
passivation layer to avoid surface states
(see Sec.~III C and Appendix~C), 
we show in Fig.~\ref{fig:05}(c,d)
that the two topmost states are indeed $P$-like and $S$-like,
bearing resemblance to the $\bm{k{\cdot}p}$ model results. The
QD considered consists of total of 512 atomic sites, including
the passivation layer atoms shown by smaller red dots in
Fig.~\ref{fig:05}(a). The QD energy levels are shown together with the
bulk band structure in Fig.~\ref{fig:05}(b) and we can see that the
$P$-like state lies above the $S$-like state. We defer the
detailed description of our Ge QD model to Sec.~III C and proceed to
analyze simplified TB models that also show the reversal of energy level ordering.

\subsection{Square lattice} 
\begin{figure}[tbp]
\begin{centering}
\includegraphics[scale=0.37]{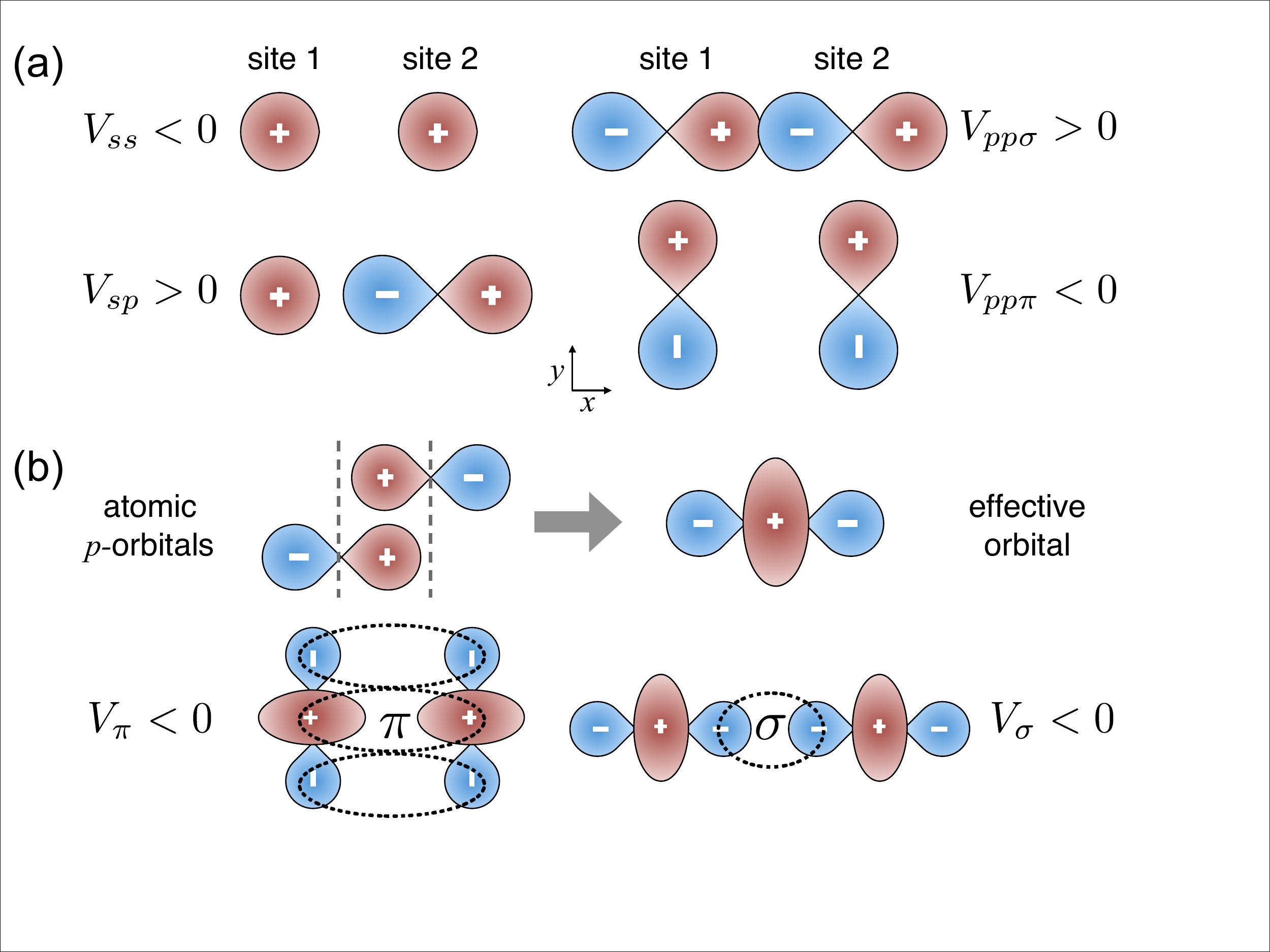} 
\par\end{centering}
\caption{Overlap (hopping) parameters of atomic and effective orbitals.
(a) Schematic representation of  the interatomic hopping parameters.
(b) Two adjacent atomic $p$-orbitals in zinc-blende or diamond lattice
can be approximated as an effective orbital. In such cases,
$\pi/\sigma$ bonds may have different signs of hopping parameters 
compared to simple cases described in (a).}
\label{fig:06}
\end{figure}

A square lattice provides a simple TB model that still contains the reversed level ordering. The transparency of this approach allows us to develop an intuitive understanding of the origin of the reversed level ordering and to further support it using an effective Hamiltonian for a continuum model.
We consider a square lattice of $N$ by $N$ atomic sites and express all
parameters of the ensuing $3N^2$-dimensional Hamiltonian matrix in
terms of 
\begin{equation}
E_{sp}\equiv E_s-E_p > 0.
\end{equation}
$p_z$-orbitals are assumed to be decoupled from the other orbitals 
and are disregarded.
By choosing $V_{ss}/E_{sp}=-0.01$,
$V_{pp\sigma}/E_{sp}=0.01$, $V_{pp\pi}/E_{sp}=0.01$, and $V_{sp}\ll
E_{sp}$, we obtain bulk band structure of this 2D crystal that is
analogous to that of a 3D zinc-blende semiconductor. The equal sign of
the $\sigma$- and $\pi$-bond hopping parameters can be understood from
the analogous situation of effective $p_x$-, $p_y$-orbitals described in 
Fig.~\ref{fig:06}(b).
These effective orbitals can be considered as basis for the VB states 
that comprise the bonding $p$-orbitals, 
while conduction band (CB) states arise from the antibonding $s$-orbitals.
However, for clarity of the description of our model,
we keep the representations of simple atomic orbitals shown in Fig.~\ref{fig:06}(a)
without losing generality of our results.

\begin{figure}[htbp]
\includegraphics[scale=0.3]{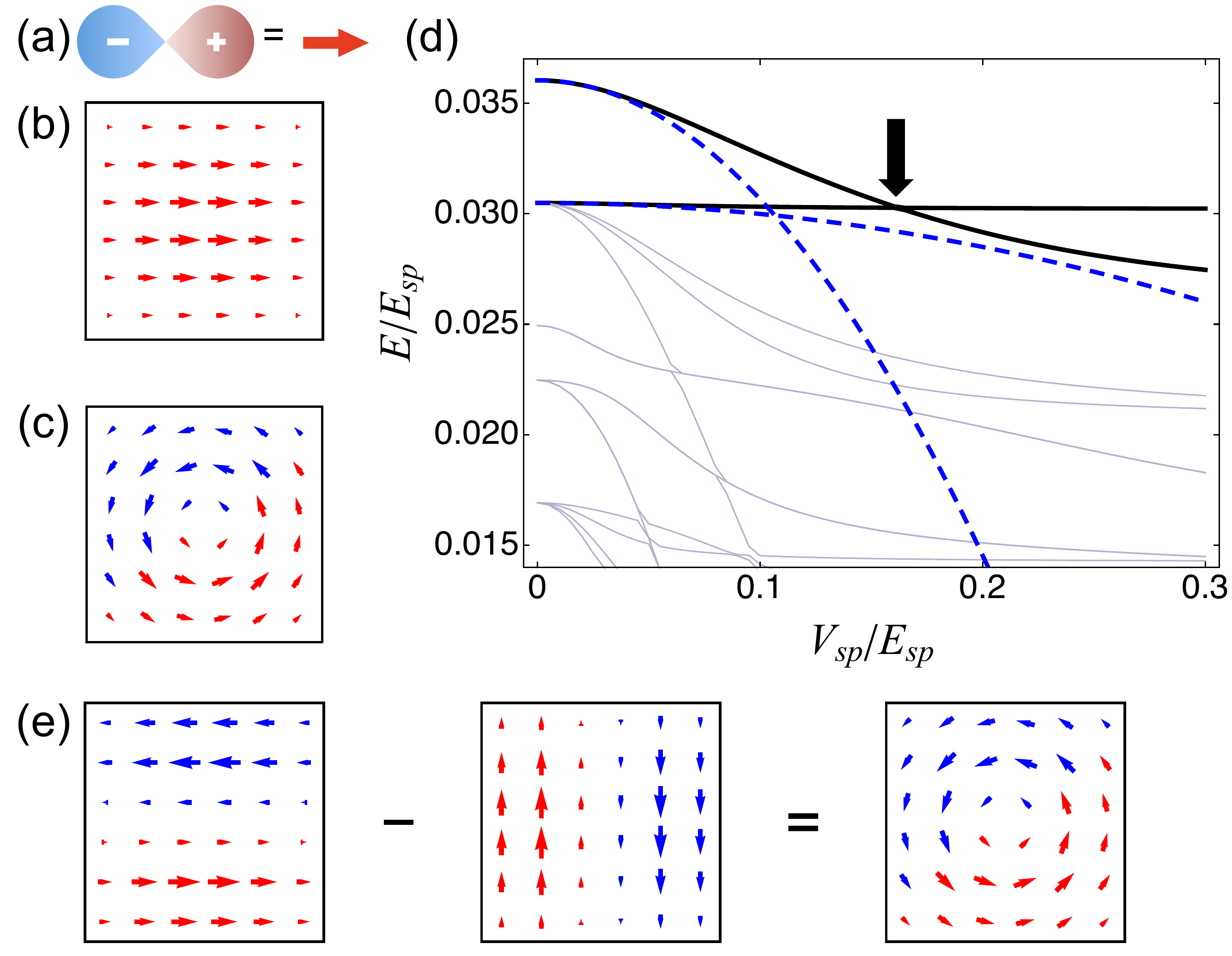}
\caption{Square QD model with 36 atoms.
(a) Vector (arrow) representation of atomic $p$-orbitals. 
The length  is the total amplitude and the direction indicates 
the composition of $p_{x}$- and $p_{y}$-orbitals.
(b)$S$-like and (c) $P$-like vortex hole states are
shown for $V_{sp}=0$ using the arrow representation.
(d) VB energy levels as a function of $V_{sp}$. For $V_{sp}=0$ the ground state is $S$-like. 
At a larger $V_{sp}$ the ordering is reversed, marked by a thick arrow, 
and the ground state becomes the $P$-like vortex state.
Dashed lines are 
analytical perturbative results 
(see text). 
For comparison with the analytical result, 
$V_{ss} \to 0$ limit is considered for both.
(e) Schematic diagram that explains how vortex state $(12)_{x}-(21)_{y}$ 
forms from two $P$-like states $(12)_{x}$ and $(21)_{y}$ (see text). 
}\label{fig:07}
\end{figure}

Turning our attention to confined systems, we look for the
eigenfunctions in the form of linear combination of atomic orbitals,
\begin{equation}\label{eq:07a}
|\psi\rangle = \sum_{{\bm R},\alpha}
c_{\alpha}({\bm R}) |\alpha({\bm R})\rangle,
\end{equation}
where $|\alpha({\bm R})\rangle$ is the
$\alpha\in\{s,p_x,p_y\}$-orbital. The coefficients $c_{\alpha}({\bm R})$ can
be understood as the multi-component envelope function and
$\sum_\alpha |c_{\alpha}({\bm R})|^2$ is the probability of an
electron being at the site ${\bm R}$ in the given state. Examples of this
probability in the $S$-like and $P$-like states in 3D diamond lattice
system are shown in Fig.~\ref{fig:05}(c,d). To analyze the
wavefunctions on a 2D square lattice, it is useful to introduce the
following basis
\begin{equation}
(n_{x}n_{y})_{\alpha}\equiv\frac{2}{N+1}
   \sum_{{\bm R}}\sin\frac{\pi n_{x}R_{x}}{(N+1)a}
\sin\frac{\pi n_{y}R_{y}}{(N+1)a}|\alpha({\bm R})\rangle,
\label{eq:03}
\end{equation}
where the positive integers $n_{x}$ and $n_{y}$ can be understood as
the quantum numbers for the envelope functions,
${\bm R}=(R_x,R_y)$ corresponds to lattice point position and $a$
is the lattice constant. For $V_{sp}=0$, the TB Hamiltonian in Eq.~(\ref{eq:05}) can be
diagonalized exactly, yielding the usual sequence of CB states,
starting with $(11)_s$, $\{(12)_s,(21)_s\}$, $(22)_s$, and $\{(31)_s,(13)_s\}$.
Here, using the notation from Eq.~(\ref{eq:03}),
the states are ordered by increasing energy, and $\{\ldots\}$ indicate
degenerate states. For an increasing hole energy, going down from the top of the VB,
the sequence begins with $\{(11)_x,(11)_y\}$ and $\{(12)_x,(12)_y,(21)_x,(21)_y\}$, i.e., 
a conventional ordering of energy levels with the $S$-like hole ground state
(twofold degenerate) and the $P$-like first excited states (fourfold degenerate).
There is no coupling between VB and CB states in this simple case.

In Fig.~\ref{fig:07} we show the evolution of the energy levels with the hopping parameter
$V_{sp}$ for a square lattice with 36 atoms. Once $s$-$p$ hybridization is included
($V_{sp}\neq 0$), the VB and CB states start to couple with each other
leading to level repulsion and removal of some degeneracies. As the $V_{sp}$
increases, $S$-like states are pushed down by CB levels and $P$-like states
split into four levels (with $V_{pp\pi}=V_{pp\sigma}$). For a sufficiently large hopping parameter, 
$V_{sp}\ge V_{sp}^{\rm{cross}}\approx0.17E_{sp}$, one of these $P$-like states 
crosses the $S$-like $\{(11)_x,(11)_y\}$
manifold\cite{note1} and becomes the hole ground state.
Remarkably, over the range of displayed values, this state is nearly independent
of $V_{sp}$ in stark contrast with all the other states shown in Fig.~\ref{fig:07}(d).
These numerical results are further corroborated by the perturbation theory
for the two lowest states at $V_{sp}=0$ (dashed lines).

\begin{figure}[tbp]
\begin{centering}
\includegraphics[scale=0.35]{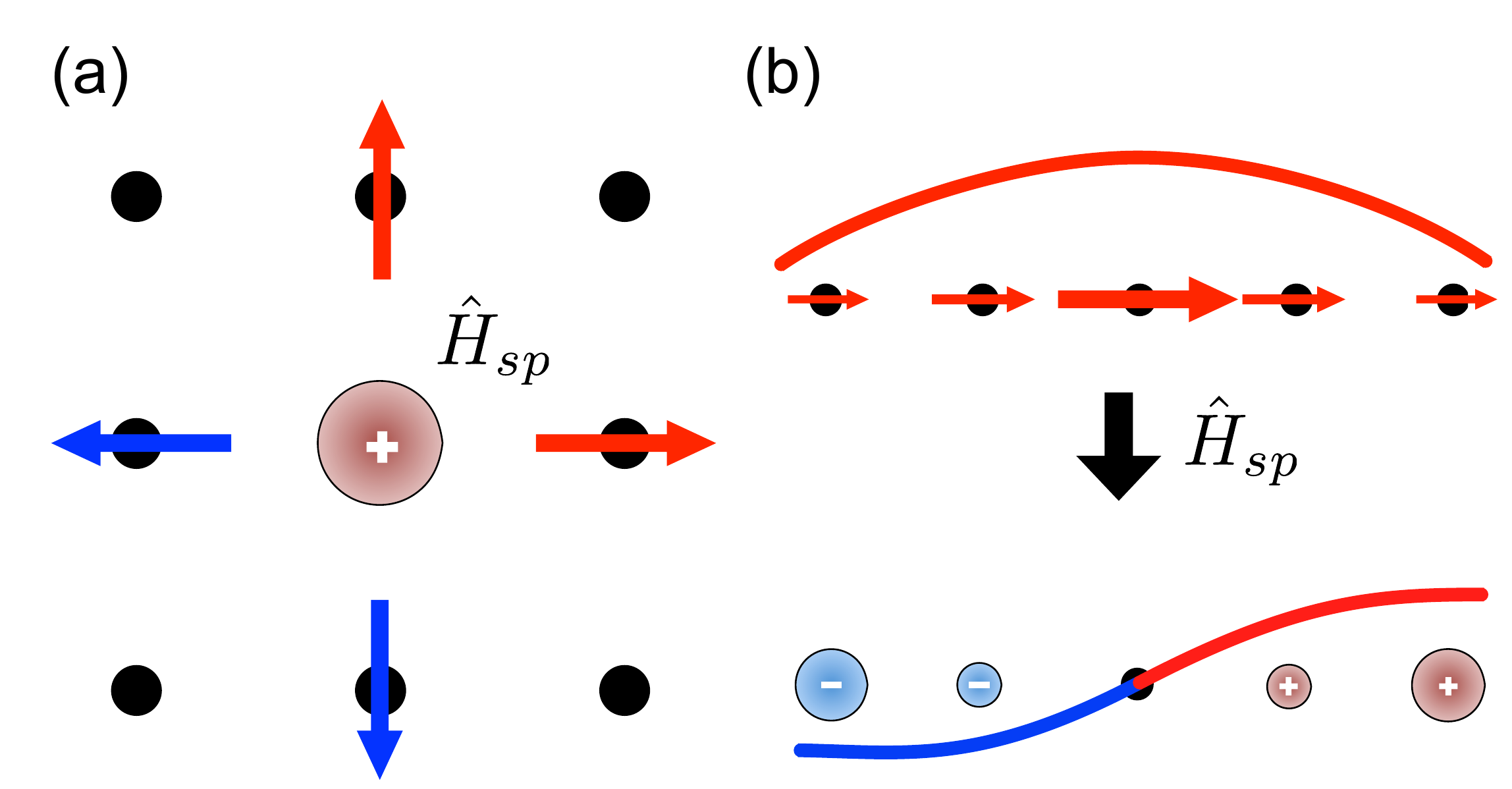}
\par\end{centering}
\caption{
Illustrations of how $\hat{H}_{sp}$ interaction effectively behaves like divergence operator 
to the envelope function 
[see Eq.~(\ref{eq:Hsp})].
(a) $\hat{H}_{sp}$ acting on an $s$-orbital
center results in four $p$-orbitals at the neighboring sites.
This is analogous to divergence of a vector field in continuum approximation.
(b) $\hat{H}_{sp}$ acting on $p$-orbitals with varying amplitude in 1D chain 
gives $s$-orbitals with the amplitudes given by the derivative of $p$-orbital amplitudes.}
\label{fig:09}
\end{figure}

What is the origin of the $P$-like hole state insensitivity to $V_{sp}$?
How can we interpret an unconventional ordering of the energy levels at the
top of the VB? To elucidate this puzzling behavior we introduce a simple
vector representation for an individual atomic $p$-orbitals, shown in Fig.~\ref{fig:07}(a).
We assign an arrow to represent $(c_{p_x}({\bm R}),c_{p_y}({\bm R}))$ as a vector,
recalling Eq.~(\ref{eq:07a}).
In this representation we can intuitively associate the VB level ordering
with the formation of corresponding orbital textures.
One of the $S$-like states is described by the orbital texture 
shown in Fig.~\ref{fig:07}(b), the other has the same texture
rotated by $90^\circ$. The $P$-like state shown in Fig.~\ref{fig:07}(c), on the 
other hand, exhibits a vortex in the orbital texture.
This state, orbital vortex state (OVS), 
is distinguished among four degenerate $P$-like states at $V_{sp}=0$.
It consists of $P$-like states with collinear wavefunctions as
shown in Fig.~\ref{fig:07}(e), indicating that the OVS wavefunction
is approximatelly $(21)_y-(12)_x$ up to normalization. Analysis of the
actual wavefunction (besides the orbital textures), confirms this 
conclusion as we describe below. We now derive a continuum
approximation of our TB model which explains why the vortex state 
in the orbital texture is less sensitive to the $s$-$p$
hybridization than the $S$-like state in Fig.~\ref{fig:07}(b).

We focus on the states at the top of the VB and take the
$s$-$p$ hybridization as a weak perturbation, the small parameter being
$V_{sp}/E_{sp}$. The TB Hamiltonian downfolded into the
subspace of $p_x$-, $p_y$-orbitals reads
\begin{equation}
\begin{split}
\hat{H}_{\rm eff}=\hat{H}_{p}+\Delta\hat{V}
=\hat{H}_{p}-\frac{1}{E_{sp}}\hat{P}_p\hat{H}_{sp}^2\hat{P}_p.
\end{split}\label{eq:DeltaV}
\end{equation}
The projection operator to the $p_x$, $p_y$ subspace is denoted by
$\hat{P}_p$ and $\hat{H}_{sp}$ is the part of the TB Hamiltonian
proportional to $V_{sp}$. We also assume that the VB and CB band
widths are small compared to $E_{sp}$. We write $\hat{H}_{sp}$ as
\begin{equation}
\hat{H}_{sp}=V_{sp}\sum_{{\bm R},{\bm
\delta},i=x,y}[\cos(\theta_{i{\bm \delta}})|s({\bm R})\rangle\langle
p_i({\bm R}+{\bm \delta})| + {\rm h.c.}],\label{eq:Hsp}
\end{equation} 
with ${\bm R}$ the lattice sites, $\bm\delta$ the nearest neighbor
vectors, $i=x,y$ index for $p_x,p_y$-orbitals, and
${\rm h.c.}$ for Hermitian conjugate.
The direction cosine of ${\bm \delta}$ is written as $\cos\theta_{i{\bm\delta}}=
\hat{\bm e}_i\cdot({\bm\delta}/|{\bm\delta}|)$
in accordance with Slater-Koster rules (see Table \ref{tab:03}).
Equation~(\ref{eq:Hsp}) is visualized in Fig.~\ref{fig:09}(a).
Due to the direction cosine $\cos\theta_{i{\bm\delta}}$, hopping from a site 
${\bm R}$ to adjacent sites ${\bm R\pm \bm \delta}$ gives opposite signs to
the resultant atomic orbitals, and this can be approximated as a derivative 
operator to the envelope function in the continuum limit.
Applying the general form of a wavefunction containing only $p_x$- and 
$p_y$-orbitals
\begin{equation}\label{eq-01}
|\Psi_p\rangle = \sum_{\bm R}\sum_{i=x,y}\psi_i({\bm R})
|p_i({\bm R})\rangle,
\end{equation}
to $\Delta \hat{V}$ defined by Eq.~(\ref{eq:DeltaV}), we obtain (see Appendix A for details)
\begin{eqnarray}
&&\langle p_x({\bm R})|\Delta \hat{V}|\Psi_p\rangle
=\frac{V_{sp}^2}{E_{sp}} \nonumber \\
&&\times \Big[\psi_x({\bm R}+2a\hat{\bm x})
+\psi_x({\bm R}-2a\hat{\bm x})-2\psi_x({\bm R})\nonumber \\
&&+\psi_y({\bm R}+a\hat{\bm x}+a\hat{\bm y})
-\psi_y({\bm R}-a\hat{\bm x}+a\hat{\bm y}) \nonumber \\
&&-\psi_y({\bm R}+a\hat{\bm x}-a\hat{\bm y})
+\psi_y({\bm R}-a\hat{\bm x}-a\hat{\bm y}) \Big].\label{eq:12}
\end{eqnarray}
Assuming slow variations of the wavefunction at the atomic length scale, we can
make the continuum approximation,
\begin{equation}
\begin{split}
\langle p_x({\bm R})|\Delta \hat{V}|\Psi_p\rangle
&\approx
\frac{4V_{sp}^2a^2}{E_{sp}}\left[\frac{\partial^2\psi_x}{\partial
x^2}({\bm R})+\frac{\partial^2\psi_y}{\partial 
x\partial y}({\bm R})
\right] \\
&=\frac{4(V_{sp}a)^2}{E_{sp}}\frac{\partial}{\partial 
x}\nabla\cdot\bm{\psi}({\bm R}).
\end{split}\label{eq:13}
\end{equation}
We represent the wavefunction from Eq.~(\ref{eq-01}) as a vector field
${\bm \psi}({\bm R})=[\psi_x({\bm R}),\psi_y({\bm R})]^T$.
The two components in ${\bm \psi}({\bm R})$ refer to
wavefunctions in the $p$-orbital space and its divergence
term is defined as
$\nabla\cdot\bm{\psi}=\partial_x\psi_x+\partial_y\psi_y$.
An equation analogous to Eq.~(\ref{eq:13}) can be found for $\langle p_y({\bm R})|\Delta
\hat{V}|\Psi_p\rangle$.

By replacing discrete lattice points ${\bm R}$ 
by an integral over ${\bm r}$, we arrive at
$H_\psi=\langle\Psi_p|\hat{H}_{\rm eff}|\Psi_p\rangle$, 
the continuum Hamiltonian projected to the
$p$-orbital subspace,
\begin{equation}
\begin{split}
&H_\psi=
\int d^2{\bm r}\\
&\times\left[
V_{p\sigma}a^2\bm{\psi}^*({\bm r})\nabla^2\bm{\psi}({\bm
r})
-\frac{4(V_{sp}a)^2}{E_{sp}}
|\nabla\cdot\bm{\psi}({\bm r})|^2
\right].
\end{split}\label{eq:continuum}
\end{equation}
The first term is the usual TB kinetic energy that accounts only for
the $p$-orbital part of the wavefunction and the second term reflects
the effect of $s$-orbital admixtures. 
It is a negative (positive in hole representation) 
divergence term that penalizes certain orbital textures.
For a texture as in Fig. 6(b), where all vectors are mutually parallel,
the zero boundary conditions imply that vector lengths must change along
the `streamlines' causing a large value of $|\partial_x \psi_x(r)|$.
On the other hand, the vortex orbital texture in
Fig.~\ref{fig:07}(c) allows to combine a small value of the divergence
term in Eq.~(\ref{eq:continuum}) with the boundary conditions satisfied.
At the same time, this texture requires that there be a node in the
center of the QD so that divergent kinetic energy (first term) in
Eq.~(\ref{eq:continuum}) is avoided. 
Hence the OVS becomes the topmost state in the VB at large
enough values of $V_{sp}$. By construction, the divergence term only
occurs with multicomponent wavefunctions that can be represented by a
vector field (orbital texture) while single-component wavefunctions
such as the CB states have the usual $S$-like ground
state. Details on the continuum model are given in Appendix~A.

This intuitive understanding of $V_{sp}/E_{sp}$ insensitivity of OVS
based on continuum limit can be confirmed in the discrete square lattice TB model.
Also in second order perturbative calculation 
with respect to $V_{sp}/E_{sp}$,
analytical expressions can be obtained for the discrete lattice, 
and we will show below with Eq.~(\ref{eq:08}) that both results 
precisely match.
Results of this calculation are shown by the dashed lines
in Fig.~\ref{fig:07}(d). Although we do not expect 
the perturbative calculation to give quantitatively precise value of
$V_{sp}^{\rm cross}$, the difference from the full
(non-perturbative) TB calculation is not very large.
The unperturbed wavefunctions of the $S$-like states are
\begin{equation}\label{eq:13tmp}
(11)_{x},\,(11)_y
\end{equation}
and the four degenerate ($V_{sp}=0$) $P$-like states are 
\begin{equation}\begin{split}\label{neq:13}
(12)_{x}-(21)_{y}, (12)_{x}+(21)_{y}, \\
(21)_{x}-(12)_{y}, (21)_{x}+(12)_{y},
\end{split}\end{equation}
up to normalization. Since $\Delta \hat{V}$ of
Eq.~(\ref{eq:DeltaV}) is symmetric to inversion both in the $x$ and 
$y$ directions, we can straightforwardly derive coupling rules between
the VB and CB such as
\begin{equation}\label{eq-03}
\langle(11)_{s}|\Delta \hat{V}|(12)_{i}\rangle 
\begin{cases}
=0, & i=x\\
\neq0, & i=y.
\end{cases}
\end{equation}
Carrying out the
degenerate-level perturbative calculation, we find that the 
degenerate $P$-like VB levels split into four states, 
and the OVS is written as,
\begin{equation}\label{eq-02}
  |\rm{OVS}\rangle = 1/\sqrt{2}\big[|(12)_{x}\rangle-|(21)_{y}\rangle\big].
\end{equation}%
Owing to the coupling rules of Eq.~(\ref{eq-03}), we immediately find
$\langle (11)_s| \Delta \hat{V}|\rm{OVS}\rangle =0$. To show
that OVS is the most weakly coupled state of the four
$P$-like states, it is necessary to mention that 
the coupling strength decreases with high values of $n_x, n_y$ 
(see Appendix. E).
The lowest CB state to which OVS couples is 
$(42)_s-(24)_s$ and the corresponding matrix element of 
$\Delta\hat{V}$ is small for this state.
On the other hand, a relatively large value
of $\langle  (21)_s|\Delta \hat{V}|(11)_x \rangle$ leads
to the strong coupling of the $S$-like state with the CB state,
and to the band repulsion that makes the $S$-like state energy 
drop quicker than the OVS with increasing $V_{sp}$ as seen in Fig.~\ref{fig:07}(d).

In the $V_{ss}\to 0$ limit the energy level reversal can be analyzed quantitatively using
\begin{eqnarray} \label{eq:07}
\Delta E_{S}&=&-4\frac{V_{sp}^{2}}{E_{sp}}\frac{N-1}{N+1}\sin^{2}\frac{\pi}{N+1},\nonumber\\
\Delta E_{P}&=&\Delta E_{S}
\times\\&&\left[1-
\frac{1}{N^2-1}
\times
\left\{\frac{8\cos\frac{\pi}{N+1}\cos^{2}\frac{\pi/2}{N+1}}{(1+2\cos\frac{\pi}{N+1})\sin\frac{\pi}{N+1}}\right\}^{2}
\right],\nonumber
\end{eqnarray}
which denote the energy level shift of the $S$-like states and the
OVS, respectively. Note that the complete expression for energy is 
$E_{S}^0+\Delta E_S$ and $E_{P}^0+\Delta E_P$, where $E_{S}^0>E_{P}^0$ are 
the energies at $V_{sp}=0$.
In the limit of an infinite number of atoms
we get
\begin{equation}
\lim_{N\to\infty} \Delta E_{P}/\Delta E_{S}=
  1-\left(\frac{8}{3\pi}\right)^{2}\approx0.28,
\label{eq:08}
\end{equation}
implying that the level repulsion from the CB levels is stronger for
the $S$-like states as we have argued so far. 
Identical result to Eq.~(\ref{eq:08}) is derived within continuum model
and is shown in Eqs.~(\ref{eq:A12}) and (\ref{eq:A13}) in Appendix A.
The energy of the $S$-like state will therefore eventually drop
below the one of the OVS, provided we stay in the perturbative regime
for accordingly large values of $V_{sp}$.
Full model calculations in Fig.~\ref{fig:07}(d) show
that the crossing indeed occurs.

\begin{figure}[htbp]
\includegraphics[scale=0.5]{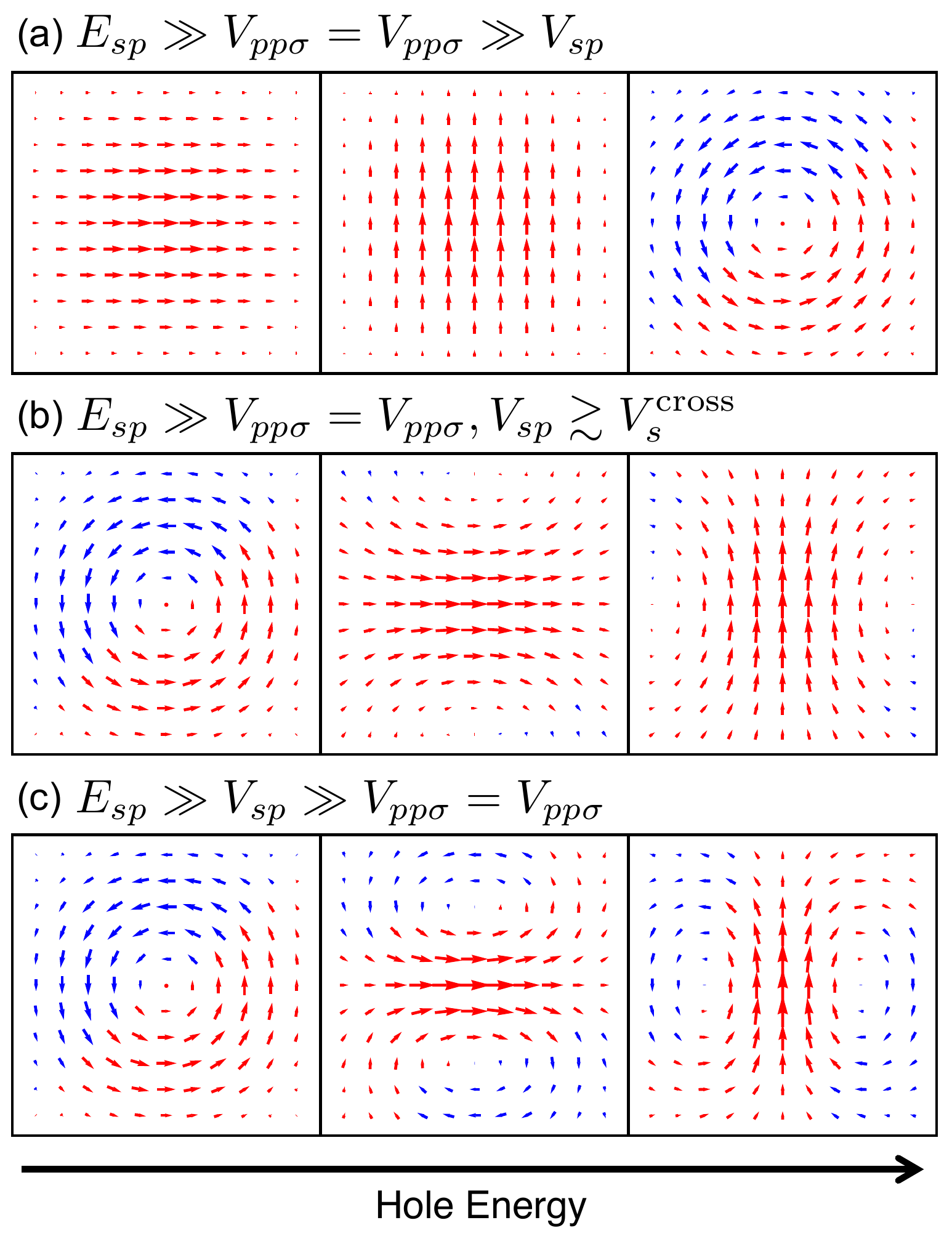}
\caption{
The evolution of orbital textures of hole states in a square lattice,
using the vector representation from  Fig.~\ref{fig:07}.
(a) In the small $V_{sp}$ limit; left panels: twofold degenerate ground
states with an $S$-like envelope, right panel: the first excited vortex $P$-like state (OVS).
(b) Orbital textures after the $S$-$P$ level crossing for
larger $V_{sp}$. The excited $S$-like states acquire curvature to reduce the
divergence term in Eq.~(\ref{eq:continuum}).
(c) The limit of $V_{sp}\gg V_{pp\sigma}$ and $V_{pp\pi}$. The $P$-like
state remains as the ground state, while the $S$-like states develop 
multi-vortex texture.}
\label{fig:08}
\end{figure}

While the reversal of the energy level ordering occurs,
$s$-$p$ hybridization promotes 
the modification of orbital textures of $S$- and $P$-like states as well.
Evolution of orbital textures as a function of $V_{sp}$ is shown 
in Fig.~\ref{fig:08}, with each row belonging to three topmost 
levels of the VB for different TB
parameter regimes, the top level states being always on the left. This figure
contains results of the full TB model calculation. In
Fig.~\ref{fig:08}(a), the states
are ordered as $(11)_x$, $(11)_y$ and OVS in the regime 
of $V_{sp}\ll E_{sp}$ before the $S$-$P$ level crossing
in Fig.~\ref{fig:07}(d).
After the $S$-$P$ level crossing occurs for larger
$V_{sp}$, the $P$-like state becomes the ground state and the
$S$-like states get appreciable admixtures from other basis states of Eq.~(\ref{eq:03})
as it can be seen in Fig.~\ref{fig:08}(b).
The original $S$-like state is gradually modified so that the
divergence of the orbital texture is decreased, i.e., the
alignment or the ``streamline'' of arrows acquires non-zero curvature as 
$V_{sp}$ increases. As $V_{sp}$ increases further
that $V_{sp}\gg V_{pp\sigma}$, $V_{pp\pi}$ in Fig.~\ref{fig:08}(c), the divergence term in
the continuum Hamiltonian of Eq.~(\ref{eq:continuum}) becomes much more
important than the $p$-$p$ 
kinetic energy, and we approach a different limit
in which the divergence of the texture becomes zero. Due to this zero
divergence, the $V_{sp}$-dependence in energy 
vanishes in the large $V_{sp}$ limit, as shown in
Fig.~\ref{fig:07}(d). The $P$-like state remains as the ground state (see Appendix
A for its explicit expression of the wavefunction), while the curvature
in the $S$-like states in (b) develops into a vortex-anti-vortex texture,
with opposite winding numbers. The
vortices are gradually introduced into $S$-like states through the boundary of the finite
sample, as we go from (b) to (c).

\subsection{Simple cubic and face centered cubic structure}

\begin{figure}[htbp]
\begin{centering}
\includegraphics[scale=0.35]{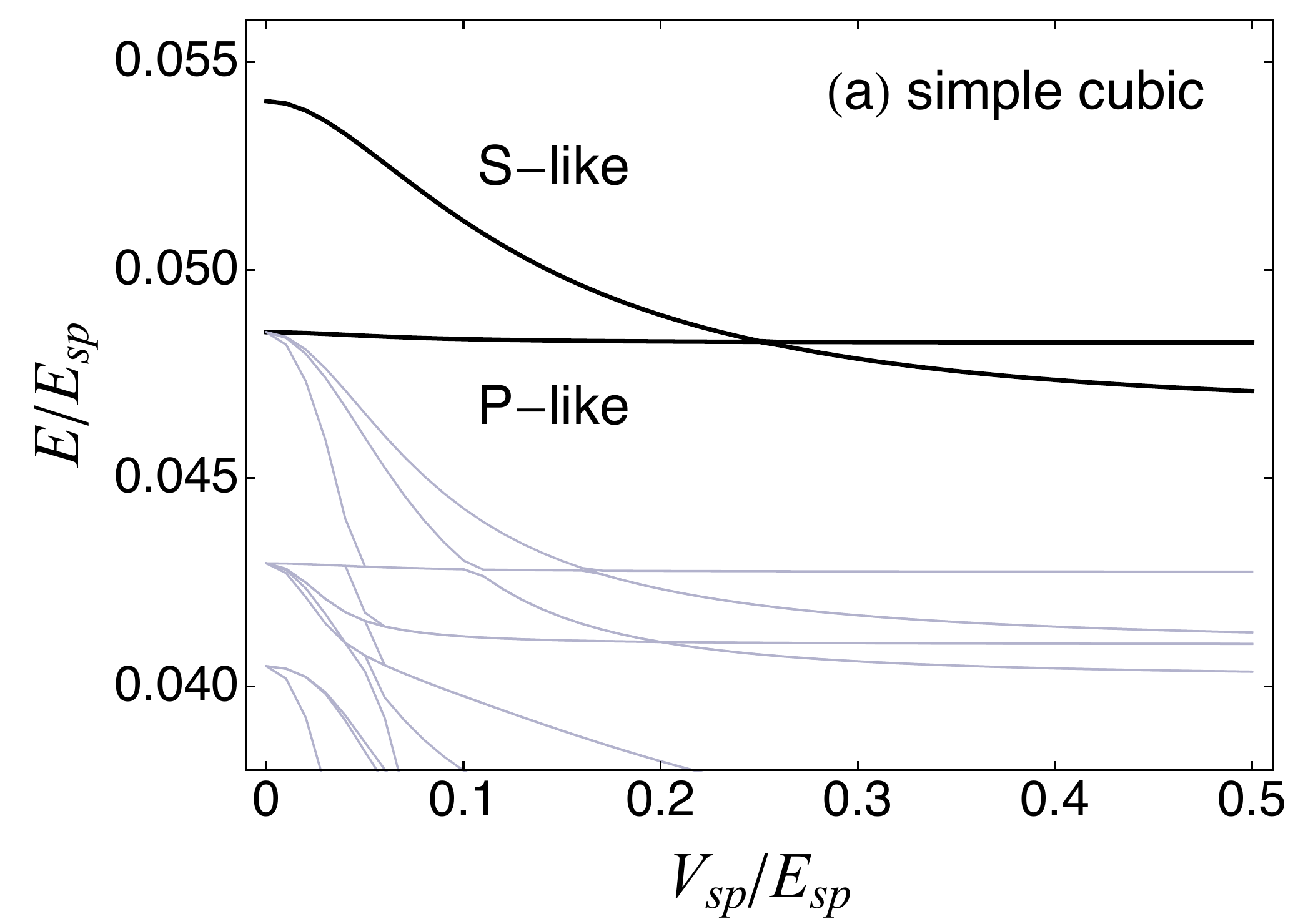}
\includegraphics[scale=0.35]{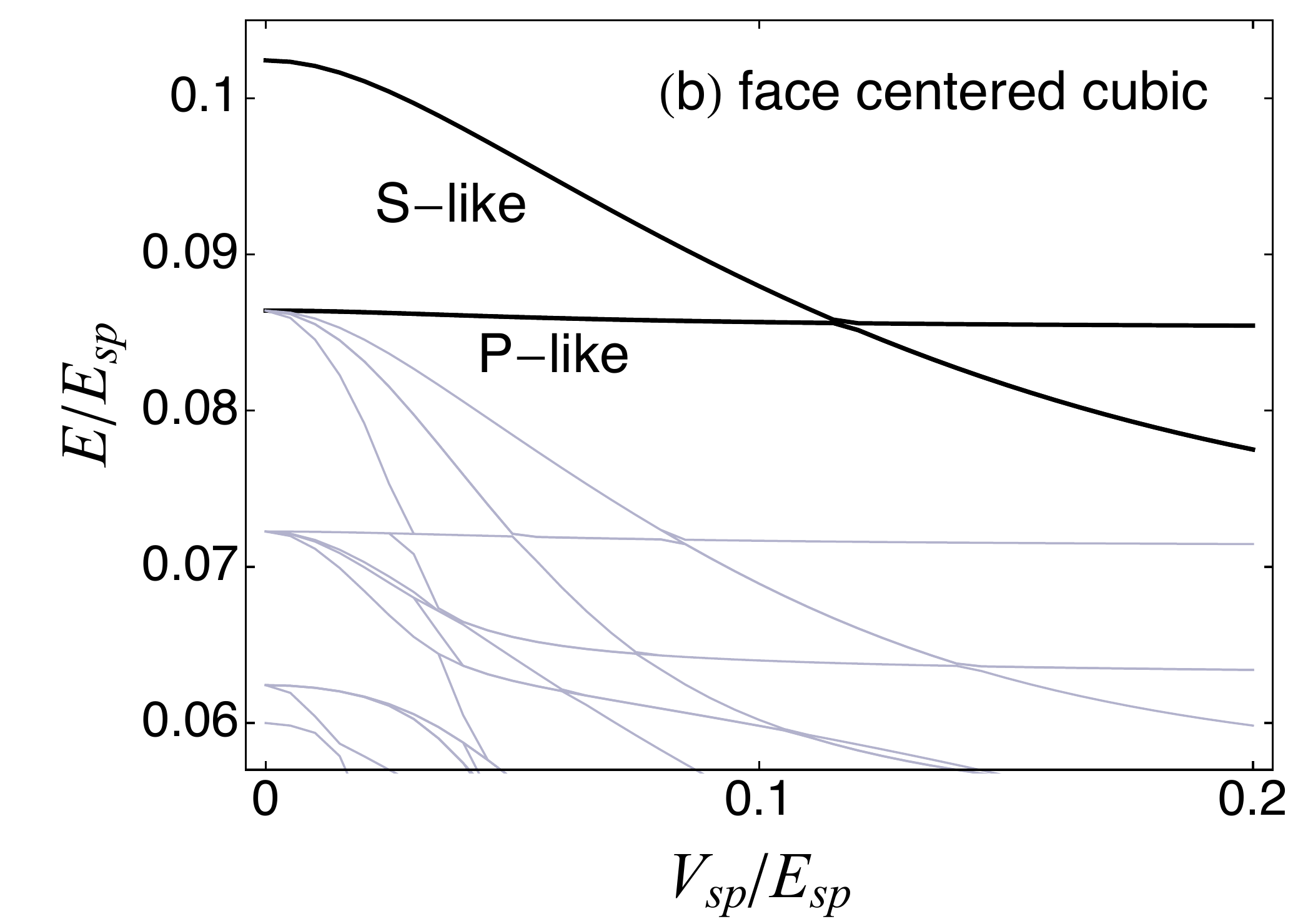}
\par\end{centering}
\caption{Energy levels in (a) SC and (b) 
FCC QDs are plotted as a function of $V_{sp}$ for fixed
$-V_{ss}/E_{sp}=V_{pp\sigma}/E_{sp}=V_{pp\pi}/E_{sp}=0.01$.}
\label{fig:10}
\end{figure}

Before we proceed to TB models of the actual semiconductor materials
(in our case, diamond structure), we consider two 3D
models as an intermediate step. 3D models include
also $p_z$-orbitals on each atomic site and the 2D basis of
Eq.~(\ref{eq:03}) can be generalized to $(n_xn_yn_z)_\alpha$,
$\alpha\in\{s,p_x,p_y,p_z\}$. The most straightforward extension of
the 2D square lattice is a 3D simple cubic (SC) lattice. Both square
and SC lattice lead to mutually analogous results.  Levels at the 
top of the VB, shown in Fig.~\ref{fig:10}(a) as a function of
$V_{sp}$, closely resemble those of Fig.~\ref{fig:07}(d) but 
the degeneracies are different. There are now three
degenerate $S$-like states, a trivial consequence of different
dimensionality, written as
\begin{eqnarray}
(111)_{x}, (111)_{y}, (111)_{z},
\end{eqnarray}
in the $V_{sp}\to0$ limit. In contrast to the square lattice, the
OVS is no longer non-degenerate and its manifold is spanned by
\begin{eqnarray}
(211)_{y}-(121)_{x},\ (121)_{z}-(112)_{y},\ (112)_{x}-(211)_{z},\quad\quad\quad
\label{eqK-03}
\end{eqnarray}
emerging from ninefold degenerate $P$-like states in the $V_{sp} \to 0$ limit.
The perturbative analysis that led
us to Eq.~(\ref{eq:07}) in the case of the square lattice can be
carried out analogously and we find, for example, the $(111)_x$ $S$-like state couples
strongly to the $(211)_s$ CB state. On the other hand, 
the OVSs couple much more weakly to the CB, owing to the high values of
$n_x$, $n_y$, $n_z$ involved just as it was the case for the square
lattice. For example, $(211)_y-(121)_x$ couples first to
no lower state than $(421)_s-(241)_s$ in the CB (see discussion in Appendix E).

The bulk four-band SC model can be downfolded into an effective
$3\times 3$ Hamiltonian as it is explained in detail in
Appendix~B. This mapping allows to link the Kohn-Luttinger parameters
of the $\bm{k{\cdot}p}$ model to the TB parameters by
comparing the band curvatures in the $\Gamma$-point and it leads to
\begin{eqnarray} \label{eq:gamma31}
\gamma_3/\gamma_1=
\frac{4V^2_{sp}/\epsilon_g}{2V_{pp\sigma}+4V_{pp\pi}+8V^2_{sp}/\epsilon_g},
\end{eqnarray}
where $\epsilon_g$ is defined in Eq.~(\ref{eq:B4}).
If we insert the parameters of the calculation in Fig.~\ref{fig:10}(a)
together with $V_{sp}=V_{sp}^{\rm{cross}}\approx 0.2 E_{sp}$, we obtain
$\gamma_3/\gamma_1\approx 0.43$. This value falls into the OVS region
in Fig.~\ref{fig:03}(a), demonstrating a correspondence between the TB
and $\bm{k{\cdot}p}$ models.

Similar results are obtained for the FCC model, the notable difference
being a somewhat smaller value of $V_{sp}^{\rm{cross}}$ which can be seen in
Fig.~\ref{fig:10}(b). The FCC model is the closest to the atomistic model
of a zinc-blende or diamond-structure material with the restriction
imposed that each primitive unit cell is described by a single effective
atomic site. The reversed level ordering at the top of the
VB clearly stems from the $s$-$p$ 
hybridization in this model and it can be intuitively understood on
the basis of the continuum model of Eq.~(\ref{eq:continuum}).
In the next subsection, we discuss the subtleties of
lifting the mentioned restriction and the move towards a more realistic model
of the QD.

\subsection{Diamond structure}

In the case of the diamond lattice (equivalent to the zinc-blende lattice
with anions replaced by the same element as cations), 
there are the antibonding
combination of the two $s$-orbitals and the bonding combinations of
$p_x$, $p_y$, $p_z$ that form the bottom of the CB and the top of the
VB, respectively, as schematically shown in Fig.~\ref{fig:01}(b).
Such an effective model was the starting point for
the calculations of the preceding subsection. The ordering of
the on-site energies with $E_{sp}>0$ is a consequence of them being
specific effective orbitals belonging to a pair of atoms rather than
actual atomic $s$- and $p$-orbitals (recall Fig.~\ref{fig:06}). From
the overall perspective of the bulk band structure implied by this
effective model, the role of $V_{sp}$ is rather minor, and
more importantly, this parameter is not responsible for the formation of the
gap between CB and VB. This role was assigned to $E_{sp}$.

The diamond or zinc-blende lattice eight-band TB models (spin is still
disregarded) is substantially different, i.e., the gap opens primarily due
to $s$-$p$ hybridization as illustrated in Fig.~\ref{fig:01}(b).
We will show that the $s$-$p$ hybridization retains
its effects that were discussed with the simpler models (see
Fig.~\ref{fig:07} and~\ref{fig:10}) also in diamond structure, 
but the value of $V_{sp}$ has to be
kept large enough so that the gap opens in the
bulk band structure. In other words, it does not make sense to consider the $V_{sp}\to 0$ limit 
in diamond structure, because the bulk band structure will not even 
have a band gap in such limit.
For confined systems, couplings between the individual bands are altered and this
$V_{sp}$-related ``gap-forming'' mechanism may fail which can manifest
itself in the formation of midgap states. Both
spurious\cite{Jaskolski2001:Vac} and physical\cite{Hapala2013:PRB}
midgap states have been discussed in the literature.\cite{Winkler:2003}
The latter one can be avoided by suitable passivation of the QD
surface, for example by covering the QD surface by a layer of hydrogen
atoms.

We focus on germanium as a material with diamond lattice 
guided by the $\bm{k{\cdot}p}$ calculations in Fig.~\ref{fig:03}. 
We use the bulk TB model parameters $E_{sp}=-8.41\,\rm{eV}$,
$V_{ss}=-1.695\,\rm{eV}$, $V_{pp\sigma}=4.065\,\rm{eV}$,
$V_{pp\pi}=-1.05\,\rm{eV}$ which consider only nearest-neighbor
interactions~\cite{Chadi1975:PTSb} 
(see Appendix~B).  Realistic effective masses are
obtained for the bulk band structure with $V_{sp}=2.3\,\rm{eV}$,
but this simplified model with nearest-neighbor interactions
predicts too high CB energy at $L$-point in the Brillouin zone,
and therefore, it fails to reproduce the indirect band gap in natural germanium. 
A 344-atom (14-atomic-layer) QD of approximately cubic shape is first constructed from Ge atoms
shown as larger blue dots in Fig.~\ref{fig:05}(a).
Then, the whole structure is covered by an extra
``passivation layer'' of distinct atoms (smaller red dots in the same
figure). If this procedure is skipped, energy levels close to the bulk
band gap have wavefunctions localized at the QD surface suggesting
that the aforementioned failure of the ``gap-forming'' mechanism has
taken place. To avoid it, we adjust the passivation layer atom
on-site energy $E_{\rm pass}$ so that the surface states are well separated in
energy from the band gap. The details in QD construction and passivation 
procedure are described in Appendix~C. We
have verified that our results do not depend strongly on the value of
$E_{\rm pass}$ as long as the surface states do not approach the band gap.

\begin{figure}[htbp]
\begin{centering}
\includegraphics[scale=0.35]{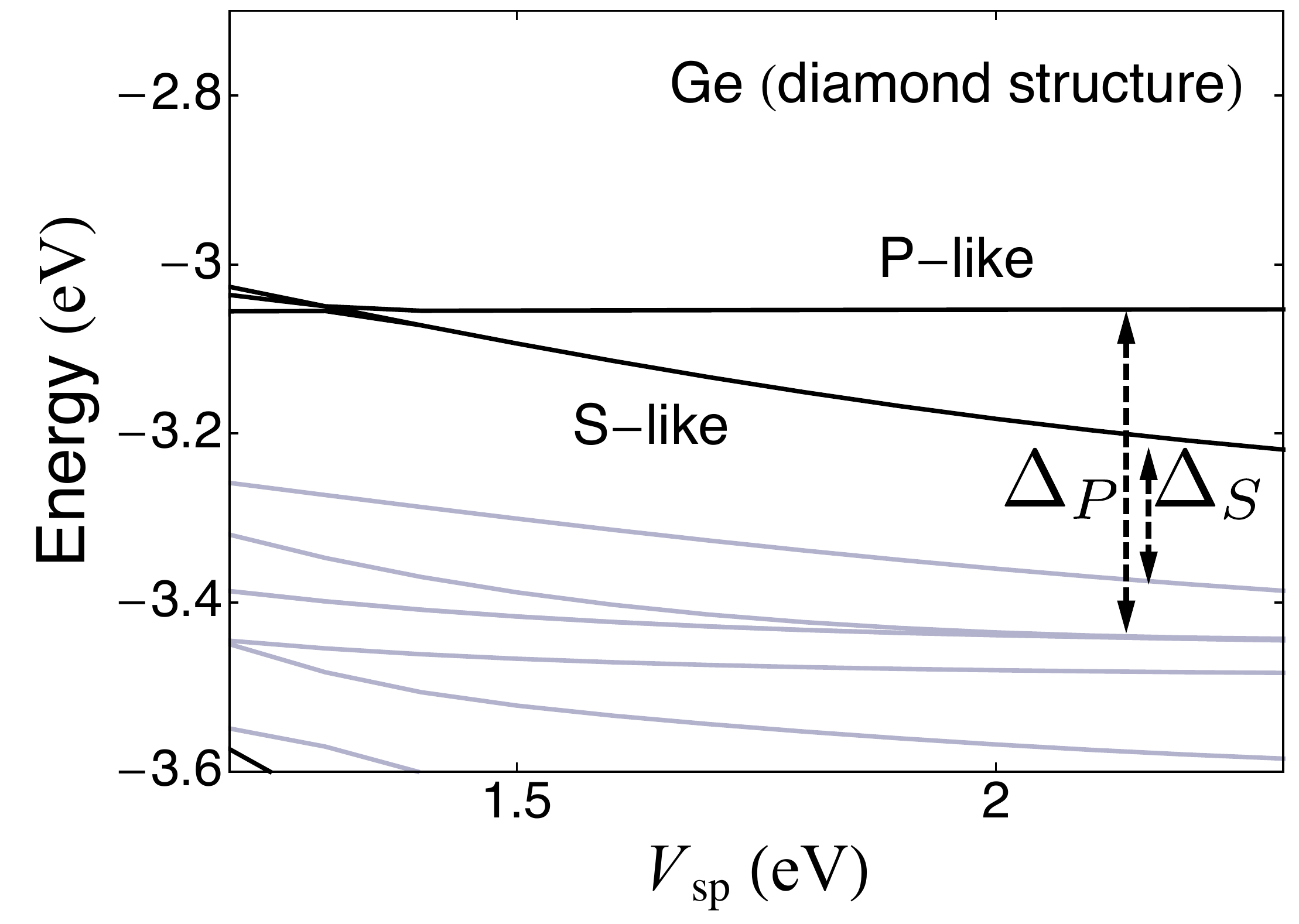}
\par\end{centering}
\caption{Energy levels at the top of the VB for a 344-atom
$\alpha$-type Ge QD covered with passivation layer as a function of 
$V_{sp}$ ($V_{sp}=2.3$~eV is the natural value for Ge). }
\label{fig:11} 
\end{figure}

Figure~\ref{fig:11} extends the calculations on which
Fig.~\ref{fig:05} was based, by considering a range of values of
$V_{sp}$. While for $V_{sp}=2.3~\rm{eV}$, the OVS occurs, we
observe that this reversed ordering of $S$-like and $P$-like hole
levels persists only down to values $V_{sp}\approx
1.2~\rm{eV}$. Character of the envelope functions, that allows to
label the two levels as $S$- and $P$-like, is demonstrated in panels
(c,d) of Fig.~\ref{fig:05}. In contrast to the SC and FCC lattice TB
models, the degeneracy of levels in Fig.~\ref{fig:11} is
lower. Rather than two triplets seen for the SC and FCC lattice, we
find a singlet and a doublet both for the $S$-like and $P$-like
states. Both magnitude and sign (which indicates the ordering of the
singlet and the doublet) of the corresponding energy splittings,
$\Delta_{S,P}$, depend on the QD size and structural
details. The latter is discussed in Appendix~C, and here we focus on
the other aspect of $\Delta_{S,P}$ that decrease with 
increasing size of the QD. Any change in the energy offset of the 
$S$- and $P$-like levels (for example, due to the changes in 
$\Delta_{S,P}$)
in Fig.~\ref{fig:11} will change the position of their crossing, 
$V_{sp}^{\rm{cross}}$. Consequently, as summarized in
Table~\ref{tab:02}, the value of $V_{sp}^{\rm{cross}}$ varies depending
on the size of the QD. We find
that regardless of the QD structure termination (two examples, $\alpha$-type
and $\beta$-type, referred to in Table~\ref{tab:02} are precisely defined 
in Appendix~C), these values tend to converge to $V_{sp}^{\rm cross}$ that is
lower than the bulk value of $2.3~\rm{eV}$. Within our
TB model of Ge, we can conclude that the OVS becomes the ground state for 
sufficiently large QDs made of this material.
In a suitably chosen material, 
our findings demonstrate that $P$-like states can be at the top of the VB.
However, for more rigorous verification,
there are other factors to be studied in more detail, such as 
next-nearest-neighbor interactions, spin-orbit interaction, and 
more elaborate surface passivation treatment.

In the $\bm{k{\cdot}p}$ calculations, effects arising from the atomistic QD structure 
that we described above, are not included.
This could raise some concerns in quantitative analyses of the level
positions in QDs. However, as long as the surface effects are not
dominant, $\bm{k{\cdot}p}$ and TB models well agree in predicting
the energy level reversal for sufficiently strong $s$-$p$ hybridization.

\begin{table}
\caption{Position of the level crossing
between the $S$-like and $P$-like states in Ge QD. The value 
$V_{sp}^{\rm{cross}}=1.30$~eV belongs to the system 
examined in Fig.~\ref{fig:11}. Size of the system is specified in
terms of the number of atomic layers considered (not counting the
passivation layer), $\alpha$ and $\beta$ refer to different 
terminations of the structure as explained in Appendix~C in detail.} 
\begin{tabular}{cc|c|c|c|c}
\hline\hline
&termination & 14 layers & 18 layers & 22 layers & 26 layers\\
\hline
$V_{sp}^{\rm{cross}}$&$\alpha$-type & 1.30 & 1.37 & 1.40 & 1.46\\
(eV)&$\beta$-type & 3.00 & 2.56 & 2.40 & 2.30\\
\hline
\end{tabular}
\label{tab:02}
\end{table}

\section{Conclusions}

In this work we provide a systematic support for the reversed level ordering of the
lowest  energy hole states, at the top of  the VB of a semiconductor QD.
In contrast to the conventional 
understanding, a nodal $P$-like state can become the ground state, having a lower energy than 
the nodeless $S$-like state. While some of the previous theoretical reports of such ``nodal 
ground states'' have been known in semiconductor  nanostructures, there was a debate if
 they are just an artifact of a  $\bm{k{\cdot}p}$ model applied to small QDs, outside 
the region of its usual validity,\cite{Bagga2006:PRB,Li2000:PRB,Persson2006:PRB,Richard1996:PRB}
and would be absent in the first-principles calculations.\cite{DoubleQD}
Similar concerns could then be raised about the experimental reports of the nodal ground state
and the reversed level ordering, if their interpretation excluding other possibilities 
also relies on the $\bm{k{\cdot}p}$ approach.\cite{Horodyska2010:PRB,Bagga2006:PRB}
Furthermore, the prior reports of the nodal ground states have yet to explain their microscopic 
origin and the underlying mechanism for their formation.

Our results for the nodal ground states in the VB are obtained from the
complementary $\bm{k{\cdot}p}$ and TB models and further supported by
a simple picture from a continuum model, thus ruling out that an unconventional 
level ordering could be limited to an artifact of  a single method. Moreover, 
the transparency of  our approach provides also a microscopic understanding 
for the evolution of the nodal states to the top of the VB and the related orbital ordering. 
We report a striking difference between the orbital textures of the nodal and 
nodeless states. In particular, we explain how the nodal ground state can be associated 
with the orbital vortex state, 
whose energy is nearly independent of the $s$-$p$ hopping parameter
over a wide range of its values.

\begin{figure}[htbp]
\includegraphics[scale=0.5]{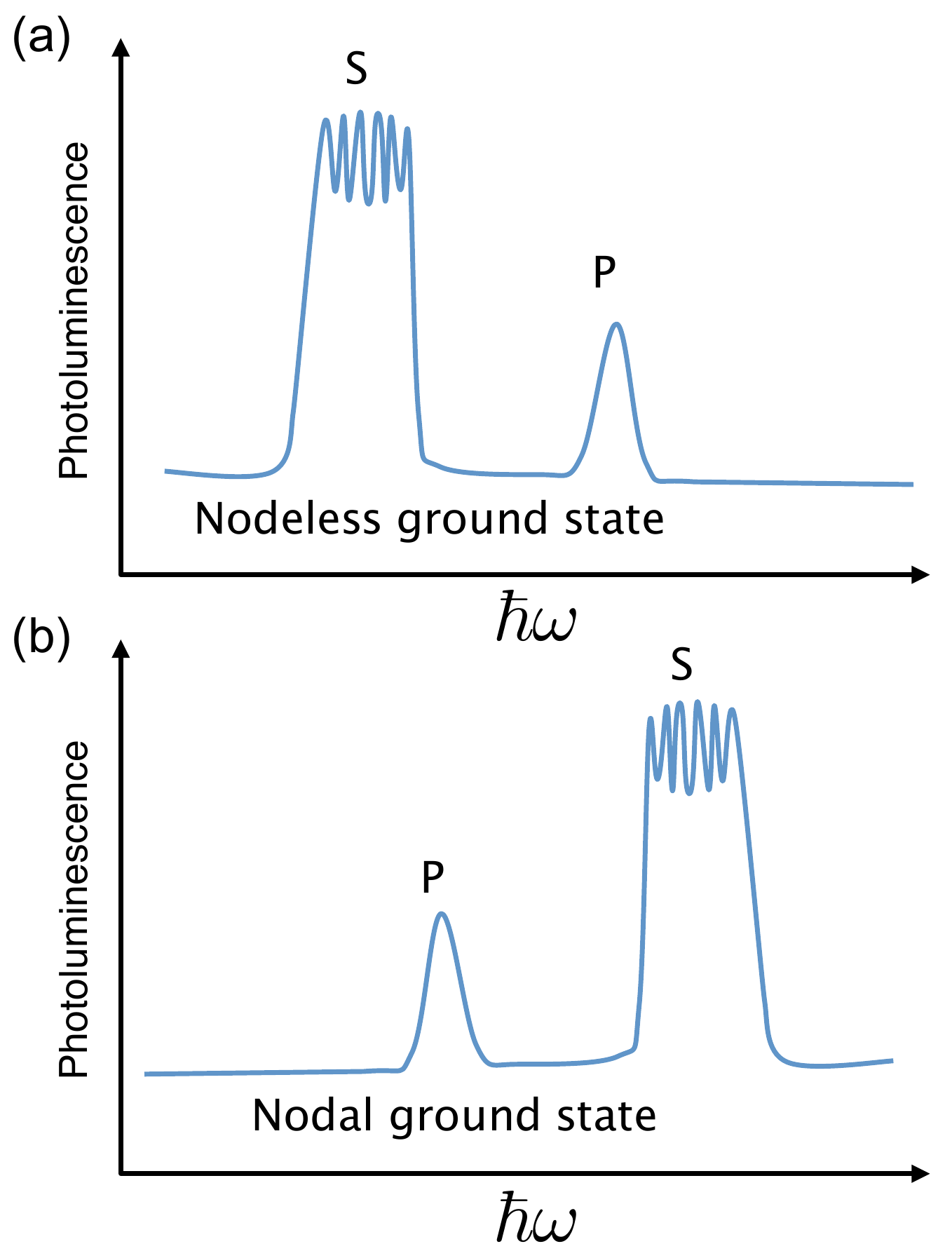}
\caption{A sketch of the photoluminescence spectrum of a QD
with a single Mn atom placed at its center. The absence (presence)
of sixfold splitting distinguishes the nodal (nodeless) state. (a) Conventional 
and (b) reversed level ordering of the hole states at the top
of the VB. The photoluminescence peaks are labeled
by the corresponding symmetry of the VB envelope functions
involved in the radiative transitions. 
The CB electron has $S$-like ground states. 
}\label{fig:PL}
\end{figure}

To test experimentally our predictions for the nodal ground state, we suggest considering 
photoluminescence measurements in QDs doped with a single magnetic 
impurity, typically Mn.  Similar pioneering experiments have already been performed
in II-VI epitaxially grown QDs.\cite{Besombes2004:PRL,Leger2005:PRL}
When spin of the single Mn atom ($S=5/2$) is decoupled from from
carriers in the QD, corresponding photoluminescence line will be
($2S+1$)-fold degenerate.
The exchange interaction of the Mn and the hole spin can be expressed 
as,\cite{Fernandez-Rossier:2011}
\begin{equation}
\hat{H}_{\rm{ex}}=J_{\rm{ex}}\hat{J}\cdot\hat{S}\delta(\bm{r}-\bm{R})
\label{eq:Hex}
\end{equation}
where $J_{\rm{ex}}$ is the hole-Mn exchange integral, 
$\hat{J}$ and $\hat{S}$ are the hole and the Mn spin 
operator with the Mn atom at position ${\bm R}$. 
This exchange interaction lifts the sixfold Mn degeneracy with
the splitting proportional to the modulus squared of the
wavefunction at the Mn position.

For a conventional nodeless $S$-like hole state, the degeneracy of the observed 
photoluminescence would
display multiple peaks in the spectrum.
Experiments are often carried out with asymmetrically shaped QDs
and in case of flat self-assembled QDs,\cite{Besombes2004:PRL} the degeneracy of
these multiplets is two, corresponding to the heavy-hole $J_z=3/2$ and
$J_z=-3/2$ components of the exciton. Each of the six doublets can be
labelled by $(J_z,S_z)$:\cite{Besombes2004:PRL,Leger2005:PRL,Fernandez-Rossier:2011}
from the ground state $(3/2,-5/2)$, $(-3/2, 5/2)$, to the highest energy state $(3/2, 5/2)$, 
$(-3/2,-5/2)$.
The measured magnitude of the $S$-like state splitting in (Cd,Mn)Te QD
was on the order of  1~meV.\cite{Besombes2004:PRL}
In contrast, if the Mn atom is located at the center of a cubic or spherical QD, 
the nodal $P$-like state should  suppress the hole-Mn exchange interaction
(as the node in an envelope function coincides with the Mn location),
producing no splitting in the photoluminescence. Schematically, these two 
outcomes are illustrated in Fig.~\ref{fig:PL} and could be used to identify
the reversed level ordering and the nodal ground
state in the VB. 
We also point out that from the theoretical point of view,
the $P$-like state should survive in asymmetrical QDs as we discuss in
Appendix D.

In an idealized $\bm{k{\cdot}p}$ 
model of a strong electron and hole confinement, 
the optical selection rules would give a 
vanishing radiative transition between the orthogonal 
$S$- and $P$-like states in the CB and VB, respectively.
This is not the case in realistic QDs of ill-defined parity. The
photoluminescence peak for the nodal state would be partially
suppressed, but not vanish, as compared to the nodeless state and indicated
in Fig.~\ref{fig:PL}.
In fact, our proposed scenario for the photoluminescence signature 
of the nodal states is providing motivation to revisit measurements of 
QDs with single Mn.\cite{Lucien} However, an appropriate identification 
would still require a detailed data analysis which we hope will be 
stimulated by our predictions for the VB orbital ordering and its implications.

Focusing on magnetically-doped QDs with many Mn impurities could also 
provide a useful test if our understanding of orbital ordering can lead to additional
means of controlling magnetism. For example, it was predicted~\cite{Abolfath2008:PRL}
that an electrostatic deformation of a quantum confinement could reversibly turn
{\em on} and {\em off} the magnetic ordering of Mn spins, even at a fixed number
of carriers. This was expected to be a consequence of the confinement-induced 
change in the carrier spin state. In contrast,  even without changing the carrier spin state,
our analysis for altering the ordering of between the nodeless and nodal states would 
imply a change in the exchange interaction (as indicated in Fig.~\ref{fig:PL}) 
and thus modify the magnetic ordering. 

A further motivation for our studies of nodal ground states in semiconductor nanostructures 
is provided by their experimental observation in other confined  systems with multiband 
electronic structure of the host lattice near the band edge.  An interesting question
would be to explore related orbital textures of such nodal states. For example, a Li donor 
impurity in Si has an ``inverted'' sequence of energy levels with a nodal ground 
state of $p$-like symmetry.\cite{Watkins1970:PRB, Jagannath1981:PRB,Pendo2013:PRB}

We thank G.~M. Sipahi for valuable discussions. 
This work was supported by the DOE-BES DE-SC0004890,
NSF-DMR 0907150, and  US ONR.

\setcounter{equation}{0}
\renewcommand{\theequation}{A-\arabic{equation}} 
\section*{Appendix A: Continuum Model}\label{Apdx:continuum}

Here we explain the derivation of our continuum model for square lattice
in Sec.~III A and the resulting effective Hamiltonian in Eq.~(\ref{eq:continuum}). 
We start from $s$- and $p$-orbitals with on-site energies $E_s$ and $E_p$
($E_s>E_p$). $s$-orbitals form the CB, and $p$-orbitals form the 
VB. We assume that the bonding within each orbital is weak and
$s$ and $p$ bandwidths are much smaller than $E_{sp}\equiv E_s-E_p$ and we can
treat the $s$-$p$ bonding (hopping) 
as perturbation. The unperturbed Hamiltonian 
in the TB model
with $s$-$p$ hopping parameter $V_{sp}=0$  can be written as
\begin{equation}\label{eq:A1}
\hat{H}_0=\sum_{\bm k}\Bigg( \epsilon_s({\bm k})|s({\bm k})\rangle\langle s({\bm k})|
+\epsilon_p({\bm k})|\psi_p({\bm k})\rangle\langle \psi_p({\bm k})| \Bigg),\quad\quad
\end{equation}
where $\epsilon_{s,p}$ are dispersion relations for $s$- %
and $p$-orbitals, respectively. Here, $|\psi_p({\bm k})\rangle$ is the Bloch
state for $p$-orbitals. The $s$-$p$ bonding as the %
perturbation is given in Eq.~(\ref{eq:Hsp}).
The bandwidths of $\epsilon_{s,p}({\bm k})$ are 
much smaller than $E_{sp}$. 
Then in the second order
perturbation theory,  through the operator %
 for the $s$-$p$ hybridization
$\hat{H}_{sp}$, the energy denominator
$[\epsilon_p({\bm k})-\epsilon_s({\bm k})]^{-1}$ can be replaced by
$(-E_{sp})^{-1}$.

To evaluate the second order term $\Delta\hat{V}$ from Eq.~(\ref{eq:DeltaV}), 
we note that in the
square lattice with a
lattice constant $a$,
we can express the effect of  $\hat{H}_{sp}$ on atomic orbitals as
\begin{equation}\begin{split}\label{eq:A2}
\hat{H}_{sp}|p_i({\bm R})\rangle & =
-V_{sp}\Bigg( |s({\bm R}+a\hat{\bm e}_i)\rangle
-|s({\bm R}-a\hat{\bm e}_i)\rangle \Bigg), \\  
\hat{H}_{sp}|s({\bm R})\rangle & = V_{sp}\sum_{i=x,y} \Bigg(
|p_i({\bm R}+a\hat{\bm e}_i)\rangle
-|p_i({\bm R}-a\hat{\bm e}_i)\rangle  \Bigg).
\end{split}
\end{equation} 

Similarly, to perform the downfolding method in Eq.~(\ref{eq:DeltaV}), 
we can also express the effect of $\hat{H}_{sp}^2$ on $p$-orbitals
\begin{widetext}
\begin{eqnarray}\label{eq:A3}
\hat{H}_{sp}^2|p_x({\bm R})\rangle & = &
-V_{sp}^2\Bigg(|p_x({\bm R}+2a\hat{\bm x})\rangle+|p_x({\bm R}-2a\hat{\bm
x})\rangle-2|p_x({\bm R})\rangle \Bigg) \\
& &-V_{sp}^2\Bigg( |p_y({\bm R}+a\hat{\bm x}+a\hat{\bm y})\rangle
-|p_y({\bm R}+a\hat{\bm x}-a\hat{\bm y})\rangle 
-|p_y({\bm R}-a\hat{\bm x}+a\hat{\bm y})\rangle
+|p_y({\bm R}-a\hat{\bm x}-a\hat{\bm y})\rangle \Bigg), \nonumber
\end{eqnarray}
and
\begin{eqnarray}\label{eq:A4}
\hat{H}_{sp}^2|p_y({\bm R})\rangle & = &
-V_{sp}^2\Bigg(|p_y({\bm R}+2a\hat{\bm y})\rangle+|p_y({\bm R}-2a\hat{\bm
y})\rangle-2|p_y({\bm R})\rangle \Bigg) \\
& &-V_{sp}^2\Bigg( |p_x({\bm R}+a\hat{\bm x}+a\hat{\bm y})\rangle
-|p_x({\bm R}+a\hat{\bm x}-a\hat{\bm y})\rangle 
-|p_x({\bm R}-a\hat{\bm x}+a\hat{\bm y})\rangle
+|p_x({\bm R}-a\hat{\bm x}-a\hat{\bm y})\rangle\Bigg).\nonumber
\end{eqnarray}
\end{widetext}
Together with Eqs.~(\ref{eq:A3}) and (\ref{eq:A4}), 
Eq.~(\ref{eq:12}) is used to derive 
the representation of $\Delta\hat{V}$ in Eq.~(\ref{eq:13}) and 
with continuum approximation, Eq.~(\ref{eq-01}) is obtained.
Now, we show how the unperturbed Hamiltonian $\hat{H}_0$
can be represented as the usual kinetic energy 
term. %
For a slowly-varying % 
wavefunction,
\begin{eqnarray}\label{eq:A5}
\hat{H}_0|p_x({\bm R})\rangle &=& V_{pp\sigma}[|p_x({\bm R}+a\hat{\bm
x})\rangle+|p_x({\bm R}-a\hat{\bm x})\rangle ] \nonumber\\
&+&V_{pp\pi}[|p_x({\bm R}+a\hat{\bm
y})\rangle+|p_x({\bm R}-a\hat{\bm y})\rangle ], \quad\quad
\end{eqnarray}
and 
\begin{eqnarray}\label{eq:A6}
\langle p_x({\bm R})|&\hat{H}_0&|\Psi_p\rangle\approx 2(V_{pp\sigma}+V_{p\pi})\nonumber\\
&+&V_{pp\sigma}a^2\frac{\partial^2\psi_x}{\partial x^2}({\bm R})
+V_{pp\pi}a^2\frac{\partial^2\psi_x}{\partial y^2}({\bm R}).\quad\quad
\end{eqnarray}
If we assume that $V_{pp\sigma}=V_{pp\pi}$, the above result
further simplifies. 
The effective Hamiltonian written for the two-component
vector-field wavefunction $\bm{\psi}_p({\bm r})$ can be derived from
$H_\psi=\langle\Psi_p|\hat{H}_{\rm eff}|\Psi_p\rangle$
and by replacing the discrete lattice points ${\bm R}$ 
by an integral over ${\bm r}$, we get
\begin{eqnarray}\label{eq:A7}
H_{\psi}&=&\sum_{i=x,y}\int d^2{\bm r} \psi_i^*({\bm r})\nonumber\\
&\times&\left[
V_{pp\sigma}a^2\nabla^2\psi_i({\bm r})
+\frac{4(V_{sp}a)^2}{E_{sp}}\partial_i
[\nabla\cdot\bm{\psi}({\bm r})]
\right],\quad\quad
\end{eqnarray}
where we have ignored the constant energy shift $V_{pp\sigma}$.
Assuming that the wavefunction is zero at the boundary due to confinement
potential, after integration-by-parts, we obtain
Eq.~(\ref{eq:continuum}),
\begin{equation}
\begin{split}\label{eq:A8}
&H_\psi=
\int d^2{\bm r}\\
&\times\left[
V_{p\sigma}a^2\bm{\psi}^*({\bm r})(\nabla^2)\bm{\psi}({\bm
r})
-\frac{4(V_{sp}a)^2}{E_{sp}}
|\nabla\cdot\bm{\psi}({\bm r})|^2
\right],
\end{split}
\end{equation}
where the first term is the usual $p$-only (TB) kinetic energy,
and the second term is the \textit{positive} divergence contribution
(in the hole representation) mediated by the
CB $s$-states. 
Using the above method, we can in principle derive an 
effective Hamiltonian for any lattice structure.

To analyze the ground state structure in the small
$V_{sp}$ limit, we start with the unperturbed kinetic
energy eigenstates, in the two-component 
notation, 
\begin{eqnarray}
&{\bm \psi}_S^0&({\bm r})=\frac{2}{L}\left[\sin\frac{\pi x}{L}\sin\frac{\pi y}{L},0\right]^T,\\ \label{eq:A9}
&{\bm \psi}_P^0&({\bm r})=\frac{\sqrt{2}}{L} 
\left[\sin\frac{\pi x}{L}\sin\frac{2 \pi y}{L}, -\sin\frac{2 \pi x}{L}\sin\frac{\pi y}{L}\right]^T,\quad\quad\quad\label{eq:A10}
\end{eqnarray}
for $S$- and OVS ($P$-like) states with their
unperturbed energies
\begin{equation}\label{eq:A11}
E_{S}^0=-\frac{2\pi^2 V_{pp\sigma}a^2}{L^2}\mbox{ and }
E_{P}^0=-\frac{5\pi^2 V_{pp\sigma}a^2}{L^2},
\end{equation}
respectively.
It is straightforward to calculate their perturbed energy in the order of
$V_{sp}^2$,
\begin{eqnarray}\label{eq:A12}
\Delta E_S&=&-\frac{4(\pi V_{sp}a)^2}{E_{sp} L^2}, \quad\quad
\end{eqnarray}
\begin{eqnarray}\label{eq:A13}
\Delta  E_P&=&\Delta E_S\left[1-\left(\frac{8}{3\pi}\right)^2\right].
\end{eqnarray}
Using $L/a=N$ and $\sin \left[ \pi/(N+1)\right] \sim\pi/N$
in the limit $N\to\infty$, this result is equivalent to Eqs.~(\ref{eq:07}) and (\ref{eq:08}).
As $V_{sp}$ increases,
the OVS energy changes slower than that of the $S$-like states, and the OVS
eventually becomes the hole ground state.

Now, we consider the opposite limit of $V_{sp}\gg V_{pp\sigma}$
and $V_{pp\pi}$. The unperturbed Hamiltonian becomes the divergence term
in Eq.~(\ref{eq:A8}) and its Schr\"odinger equation becomes
\begin{equation}\label{eq:A14}
\frac{4(V_{sp}a)^2}{E_{sp}}\nabla[\nabla\cdot\bm{\psi}(\bm{r})]=E_0\bm{\psi}(\bm{r}),
\end{equation}
with the unperturbed energy $E_0$.
With the infinite potential well limit, finite $V_{pp\sigma}$ and
$V_{pp\pi}$ require that the vector $\bm{\psi}(\bm{r})$ must vanish on
the boundary. 
Solving the differential equation 
$\nabla\cdot\psi(r)=0$ with zero boundary conditions, its solutions 
automatically satisfy the above Schr\"odinger equation with 
$E_0=0$. 
We find the ground state of the kinetic energy within
the subspace of such solutions. 
With the observation of the ground state texture, we propose a state 
in a separable form, $\psi_x(x,y)=f(x)g(y)$ and $\psi_y(x,y) = -g(x)f(y)$. 
From the zero divergence condition, we require $g(x)=f'(x)$. The new 
boundary conditions on $f(x)$, that is $f(0)=f(L)=0$ and $f'(0)=f'(L)=0$, 
can be accommodated by writing $f(x)=[h(x)]^2$ with $h(0)=h(L)=0$. 
Since the proposed form of the ground state must optimize the kinetic 
energy, we start by a quadratic polynomial $h(x)=x(L-x)$ with the least 
curvature. The form of the polynomial can be improved variationally by 
including higher orders. Therefore the proposed ground state 
wavefunction reads,
\begin{eqnarray}\label{eq:A15}
&{\bm \psi}&_P^{\infty}({\bm r})=
2C{\cdot}h(x)h(y)\big[h(x)h'(y),-h'(x)h(y)\big]^T,\quad\quad\quad
\end{eqnarray}
with the normalization constant $C$. This state 
preserves the $P$-like symmetry with the zero-divergence condition 
strictly imposed. The agreement of this wave-function and the numerically
exact solution to the continuum model, Eq.~(\ref{eq:A8}), is excellent
at the discrepancy of 1.1\% with the error defined as
$\int|\bm{\psi}_P^{\infty}(\bm{r})-\bm{\psi}_P^{\rm{num}}(\bm{r})|^2 dxdy$
for $V_{sp}\gg V_{pp\sigma}$.
Due to this
zero-divergence condition, the unperturbed energy has no $V_{sp}$
dependence and, as shown in Fig.~\ref{fig:07}(d), the energy eigenvalue
of the $P$-like state becomes independent of $V_{sp}$ as
$V_{sp}\to\infty$. The energy of the $P$-like state can be evaluated by
the expectation value of the $p$-$p$ kinetic energy term as,
\begin{equation}
E_P^{\infty}=-\frac{5.45\pi^2V_{pp\sigma}a^2}{L^2}=1.09E_P^0,
\end{equation}
in the limit of $V_{sp}\to\infty$. It is remarkable that this expression
is very close to that of the $V_{sp}=0$ limit, Eq.~(\ref{eq:A11}), which
justifies the very weak $V_{sp}$ dependence throughout for all $V_{sp}$ from the TB
result, as shown in Fig.~\ref{fig:07}(d).

\begin{widetext}

\setcounter{equation}{0}
\renewcommand{\theequation}{B-\arabic{equation}} 
\section*{Appendix B: Tight-Binding Model for Simple Cubic, Face Centered Cubic,  
and Diamond Lattice}\label{Apdx:bridge}

This Appendix presents more details on the various TB models discussed 
in the Secs.~III B and C 
and we establish a connection between the $\bm{k{\cdot}p}$ and TB models.
We first consider a bulk sample with a 
SC lattice having four orbitals,  $s$, $p_x$, $p_y$, and $p_z$. 
The relevant parameters are the difference of the 
on-site energies $E_{sp}$
and hopping parameters $V_{pp\sigma}$, $V_{pp\pi}$, $V_{ss}$ and $V_{sp}$ 
as defined in Sec.~III.
The corresponding TB Hamiltonian is given as
\begin{eqnarray}
\rm{H}^{4\times4}_{\rm{SC}}&=&\left(\begin{array}{cccc}
F(k_{x},k_{y},k_{z}) & 0 & 0 & 2iV_{sp}\sin k_{x}a\\
0 & F(k_{y},k_{x},k_{z}) & 0 & 2iV_{sp}\sin k_{y}a\\
0 & 0 & F(k_{z},k_{x},k_{y}) & 2iV_{sp}\sin k_{z}a\\
-2iV_{sp}\sin k_{x}a & -2iV_{sp}\sin k_{y}a & -2iV_{sp}\sin k_{z}a & E_{sp}+G(k_{x},k_{y},k_{z})
\end{array}\right), \label{eq:TB4}
\end{eqnarray}
where 
\begin{equation}\begin{split}
\ensuremath{F(k_{x},k_{y},k_{z})&=2V_{pp\sigma}\cos k_{x}a
+2V_{pp\pi}(\cos k_{y}a+\cos k_{z}a)},\\
G(k_{x},k_{y},k_{z})&=2V_{ss}(\cos k_{x}a+\cos k_{y}a+\cos k_{z}a),
\end{split}\end{equation}
and $a$ is the lattice constant. 
Using the L\"owdin approximation\cite{Chuang:2009}
$\rm{H}_{ij}^{3\times3}\approx\rm{H}_{ij}^{4\times4}+\sum_{\alpha\in \{s\}} H_{i\alpha}^{4\times4}\rm{H}_{\alpha j}^{4\times4}/(E-\rm{H}_{\alpha\alpha}^{4\times4})$, 
original basis space $\{s,p_x,p_y,p_x\}$ can be folded into 
the reduced Hilbert space of $\{p_x,p_y,p_x\}$. Then,
the reduced Hamiltonian is written as,
\begin{eqnarray}
H^{3\times3}_{\rm{SC}}&=&\frac{4V_{sp}^{2}}{\epsilon_g}\left(
\begin{array}{cccc}
\frac{\epsilon_g}{4V_{sp}^{2}}F(k_{x},k_{y},k_{z})+\sin^{2}k_{x}a & \sin k_{x}a\sin k_{y}a & \sin k_{z}a\sin k_{x}a\\
\sin k_{y}a\sin k_{x}a & \frac{\epsilon_g}{4V_{sp}^{2}}F(k_{y},k_{z},k_{x})+\sin^{2}k_{y}a & \sin k_{y}a\sin k_{z}a\\
\sin k_{z}a\sin k_{x}a & \sin k_{z}a\sin k_{y}a & \frac{\epsilon_g}{4V_{sp}^{2}}F(k_{z},k_{x},k_{y})+\sin^{2}k_{z}a
\end{array}
\right),\label{eq:TB3}
\end{eqnarray}
with $F_{0}=F(0,0,0)$ and $G_{0}=G(0,0,0)$, where 
\begin{equation}\label{eq:B4}
\epsilon_g=F_{0}-(E_{sp}+G_{0}).
\end{equation} 
The effective $3\times3$ Hamiltonian describes the VB only. 
Now assuming $|\bm{k}|a\ll1$, we expand each matrix element 
up to the second-order in
 $\bm{k}$ to obtain $H^{3 \times 3}$ as
\begin{eqnarray}
&\approx&\left(\begin{array}{cccc}
F_{0}+Ak_{x}^{2}+B(k_{y}^{2}+k_{z}^{2}) & Ck_{x}k_{y} & Ck_{z}k_{x}\\
Ck_{x}k_{y} & F_{0}+Ak_{y}^{2}+B(k_{z}^{2}+k_{x}^{2}) & Ck_{y}k_{z}\\
Ck_{z}k_{x} & Ck_{z}k_{y} & F_{0}+Ak_{z}^{2}+B(k_{x}^{2}+k_{y}^{2})
\end{array}\right),\label{eq:kp3}
\end{eqnarray}
where $A=-(V_{pp\sigma}-4V_{sp}^{2}/\epsilon_g)a^{2}$,
$B=-V_{pp\pi}a^{2}$, and $C=4V_{sp}^{2}a^{2}/\epsilon_g$.

On the other hand, a $6\times6$ Luttinger Hamiltonian can be transformed
into the exactly same form\cite{Luttinger} of a matrix representation as 
Eq.~(\ref{eq:kp3}) through a basis transformation
with $\gamma_{1}=-2m(A+2B)/3\hbar^{2}$,
$\gamma_{2}=-m(A-B)/3\hbar^{2}$, and $\gamma_{3}=-mC/3\hbar^{2}$.
The coefficients A, B, and C correspond to the original definition from 
Refs.~\onlinecite{Luttinger, Chow:1996}.
We provide a useful correspondence between the TB 
hopping parameters and Kohn-Luttinger parameters, 
\begin{eqnarray}
\gamma_1&=&\frac{2ma^2}{3\hbar}(V_{pp\sigma}+2V_{pp\pi}-4V_{sp}^{2}/\epsilon_g),
\nonumber\\
\gamma_2&=&\frac{ma^2}{3\hbar}(V_{pp\sigma}-V_{pp\pi}-4V_{sp}^{2}/\epsilon_g), 
\label{eq:gammas}\\
\gamma_3&=&-\frac{ma^2}{3\hbar}4V_{sp}^{2}.\nonumber
\end{eqnarray}
These expressions directly lead to Eq.~(\ref{eq:gamma31}).

We now proceed to describe the FCC and diamond lattice TB Hamiltonian matrices used
in Secs.~III B and C. The four-band FCC TB model is written as,
\begin{eqnarray}
\rm{H}^{4\times4}_{\rm{FCC}}=\left(\begin{array}{cccc}
I(k_{x},k_{y},k_{z}) & K(k_{x},k_{y}) & K(k_{x},k_{z}) & iK(k_{x},k_{y},k_{z})\\
K(k_{y},k_{x}) & I(k_{y},k_{x},k_{z}) & K(k_{y},k_{z}) & iK(k_{y},k_{z},k_{x})\\
K(k_{z},k_{x}) & K(k_{z},k_{y}) & I(k_{z},k_{x},k_{y}) & iK(k_{z},k_{x},k_{y})\\
-iK(k_{x},k_{y},k_{z}) & -iK(k_{y},k_{z},k_{x}) & -iK(k_{z},k_{x},k_{y}) & E_{0}+J(k_{x},k_{y},k_{z})
\end{array}\right)
\label{eq:FCC4}
\end{eqnarray}
where, 
\begin{eqnarray}
&&I(k_{x},k_{y},k_{z})=(V_{pp\sigma}+V_{pp\pi})(C^{+}_{xy}+C^{+}_{zx})+2V_{pp\pi}C^{+}_{yz},
\qquad\\
&&J(k_{x},k_{y},k_{z})=2V_{ss}(C^{+}_{xy}+C^{+}_{yz}+C^{+}_{zx}),\\
&&K(k_{x},k_{y})=(V_{pp\sigma}-V_{pp\pi})C^{-}_{xy},\\
&&K(k_{x},k_{y},k_{z})=\sqrt{2}V_{sp}(S^{+}_{xy}+S^{+}_{zx}),\\
&&C^{\pm}_{xy}=\cos (k_x+k_y)a/2 \pm \cos (k_x-k_y)a/2,\\
&&S^{\pm}_{xy}=\sin (k_x+k_y)a/2 \pm \sin (k_x-k_y)a/2.
\end{eqnarray}

Hamiltonian for the diamond lattice which 
contains two atoms in a primitive unit cell is written as,
\begin{equation}
\left(
\begin{array}{cc}
D_1 & M \\ M^* & D_2
\end{array}
\right),
\end{equation}
where the two diagonal blocks  $D_1=D_2={\rm diag}(E_{sp},0,0,0)$
are identical since the cation and anion atoms are the same for diamond lattice.
The off-diagonal $4 \times 4$ block $M$ can be expressed using parameters 
$V_{xx}$, $V_{xy}$, related to the parameters discussed in Sec.~III C and defined by
$V_{xx}=4(V_{pp\sigma}+2V_{pp\pi})/3$, $V_{xy}=4(V_{pp\sigma}-V_{pp\pi})/3$.
Following Chadi and Cohen,\cite{Chadi1975:PTSb}
\begin{equation}
M=\left(
\begin{array}{cccc}
4V_{ss} g_1 & -\frac{4}{\sqrt{3}}V_{sp} g_2 & 
-\frac{4}{\sqrt{3}}V_{sp} g_3 & -\frac{4}{\sqrt{3}}V_{sp} g_4\\
\frac{4}{\sqrt{3}}V_{sp} g_2 & V_{xx} g_1 & V_{xy} g_4 & V_{xy} g_3\\
\frac{4}{\sqrt{3}}V_{sp} g_3 & V_{xy} g_4 & V_{xx} g_1 &V_{xy} g_2\\
\frac{4}{\sqrt{3}}V_{sp} g_4 & V_{xy} g_3 & V_{xy} g_2 & V_{xx} g_1
\end{array}
\right),
\end{equation}
where
\begin{equation}
\begin{split}
g_1=(1/4)\{\exp[i {\bm d}_1\cdot {\bm k}]+
\exp[i {\bm d}_2\cdot {\bm k}]+
\exp[i {\bm d}_3\cdot {\bm k}]+
\exp[i {\bm d}_4\cdot {\bm k}]
\}, \\
g_2=(1/4)\{\exp[i {\bm d}_1\cdot {\bm k}]+
\exp[i {\bm d}_2\cdot {\bm k}]-
\exp[i {\bm d}_3\cdot {\bm k}]-
\exp[i {\bm d}_4\cdot {\bm k}]
\}, \\
g_3=(1/4)\{\exp[i {\bm d}_1\cdot {\bm k}]-
\exp[i {\bm d}_2\cdot {\bm k}]+
\exp[i {\bm d}_3\cdot {\bm k}]-
\exp[i {\bm d}_4\cdot {\bm k}]
\}, \\
g_4=(1/4)\{\exp[i {\bm d}_1\cdot {\bm k}]-
\exp[i {\bm d}_2\cdot {\bm k}]-
\exp[i {\bm d}_3\cdot {\bm k}]+
\exp[i {\bm d}_4\cdot {\bm k}]
\},
\end{split}
\end{equation}
with ${\bm d}_1=(1,1,1)(a/4)$, ${\bm d}_2=(1,-1,-1)(a/4)$, 
${\bm d}_3=(-1,1,-1)(a/4)$ and  ${\bm d}_4=(-1,-1,1)(a/4)$.
We use Ge parameters from Ref.~\onlinecite{Chadi1975:PTSb} for the nearest-neighbors to
keep our model simple and the corresponding bulk band structure is shown
in Fig.~\ref{fig:05}(b). 

\end{widetext}

\setcounter{equation}{0}
\renewcommand{\theequation}{C-\arabic{equation}} 
\section*{Appendix C: Issues related to the termination and
  passivation of the QD structure}\label{Apdx:passivation}

Compared to SC and FCC lattices, the diamond lattice allows for multiple ways
of termination in a finite-size structure. The choice of termination
as well as the properties of the surface layer of atoms (passivation)
often have a strong influence on the QD electronic structure. We start
by discussing the diamond lattice termination.\\
\indent A possible procedure to construct a cubic QD
is the following. We take a $3\times 3\times 3$ stack of FCC
unit cells and combine it with its copy shifted by 1/4 of the
body diagonal. The resulting 344 atoms can be divided into 14 layers
written down as ABCDABCDABCDAB in the notation of Fig.~12(a). This
procedure corresponds to the $\beta$-type termination discussed in
Sec.~III C and Table~\ref{tab:02}. The other type ($\alpha$) discussed corresponds
to DABCDABCDABCDA for 14 layers. Larger QDs are
constructed according to the same pattern: C...D (A...B) for $\alpha$
($\beta$).

\begin{figure}[htbp]
\includegraphics[scale=0.5]{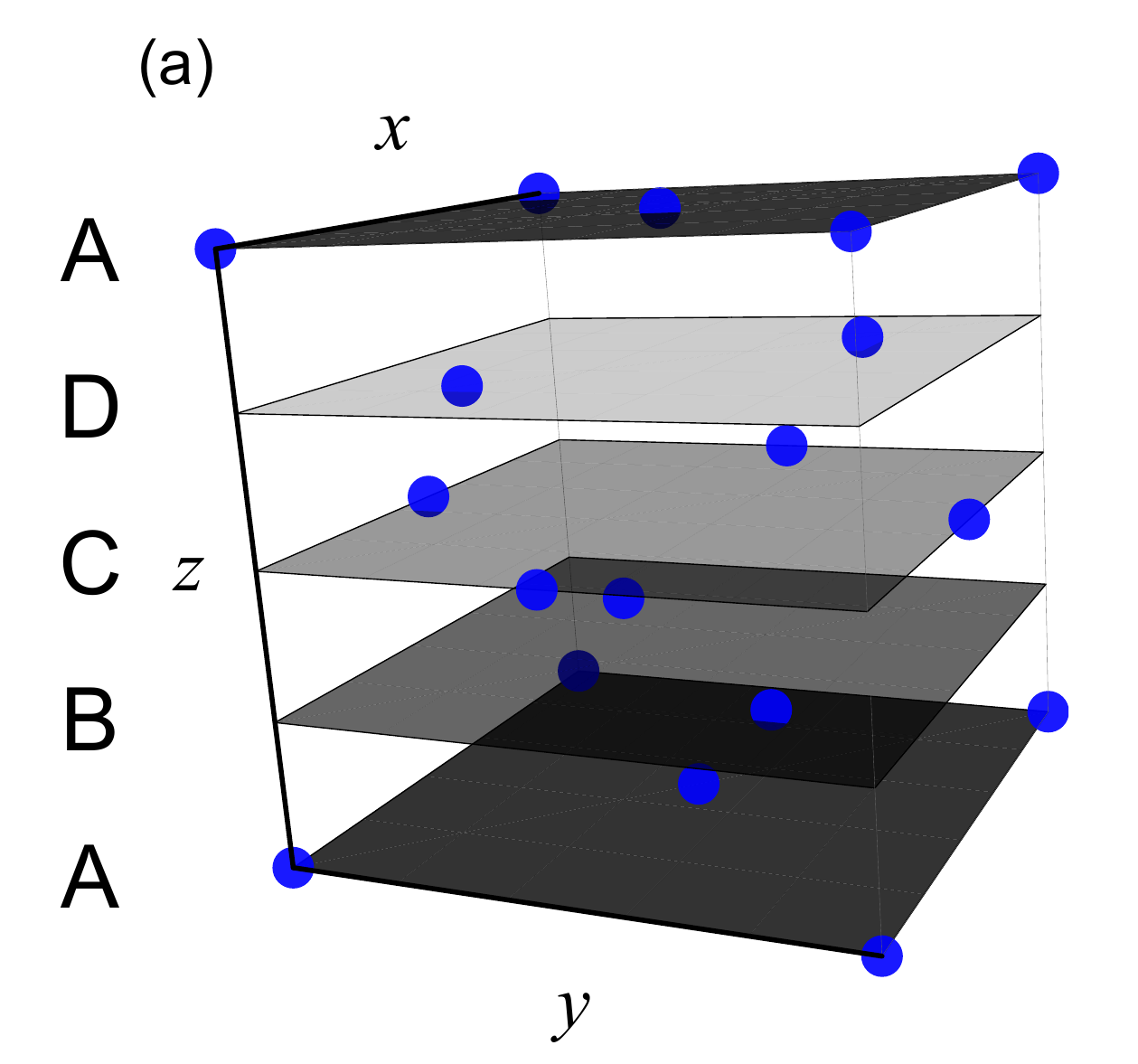}
\includegraphics[scale=0.4]{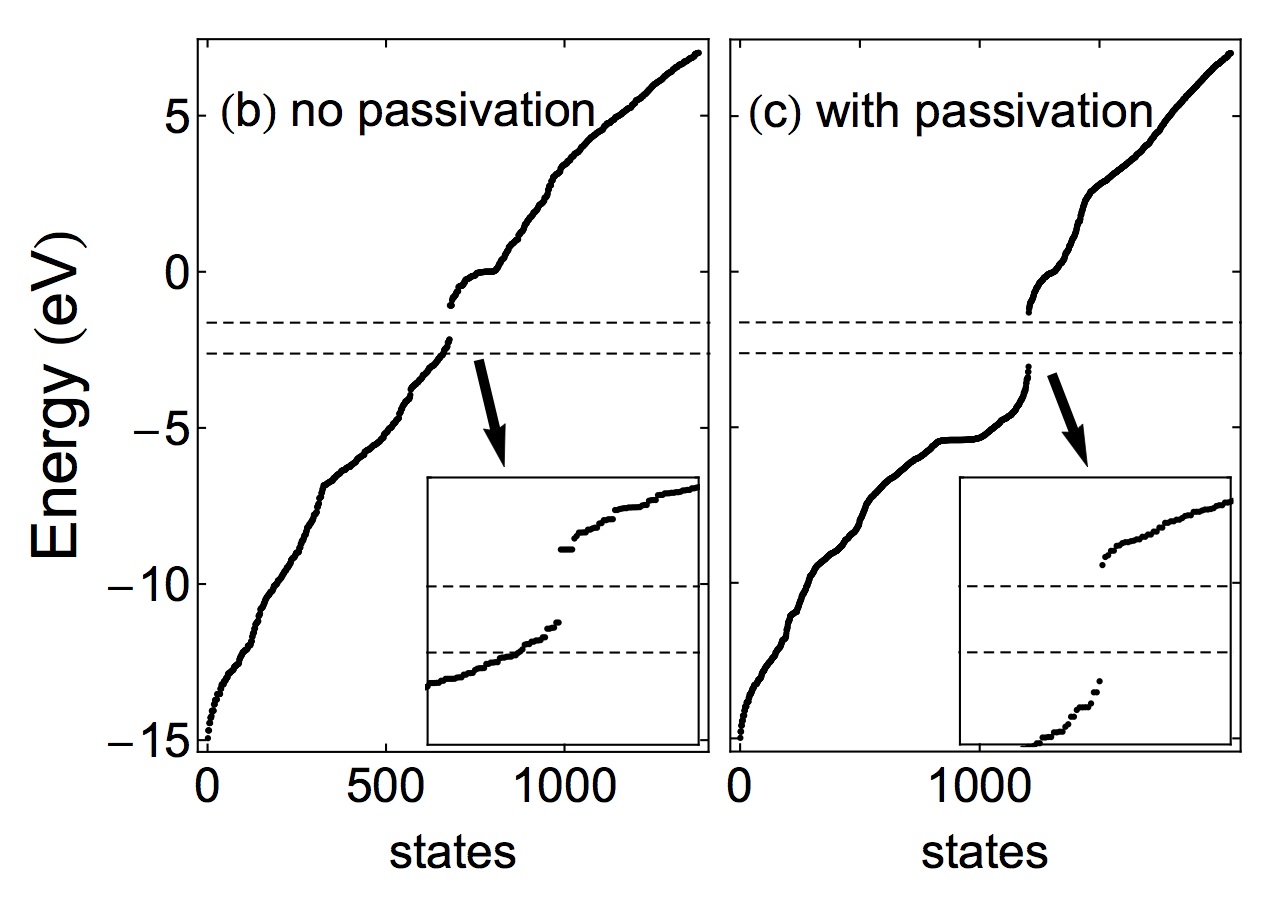}
\caption{(a) Diamond lattice unit cell with layers labeled.
Starting from the bottom face, each face at $1/4$,
$2/4$, and $3/4$ of the unit cell height are labeled as 
A, B, C, and D, respectively. The top face is equivalent to the the bottom face.
Comparison of Ge QD energy calculation (b) with and (c) without passivation 
for $\alpha$-type termination.
Insets show the states in the smaller region near the bulk band gap, denoted by 
two dashed lines. 
Note that the confinement made the gap of the QD 
larger than the bulk energy gap.
}\label{fig:C1}
\end{figure}

These two different
lattice terminations give qualitatively different QD electronic structures in
terms of the level degeneracy at the top of the VB. In the large
$V_{sp}$ limit, both $\alpha$- and $\beta$-type (passivated) clusters
show a $P$-like state at the top of the VB. However, while the state
is non-degenerate for $\beta$-type, it forms a doublet state for the
$\alpha$-type. This can be seen as a finite-crystal field
splitting of the originally three-dimensional manifold of $P$-like
states discussed in Sec.~III B for the SC and FCC lattices, a splitting
that changes sign depending on the cluster termination. Regardless of
this, an analogous plot to Fig.~\ref{fig:11} showing the
$V_{sp}$-dependence of energy levels in a $\beta$-type cluster
exhibits a crossing between $S$- and $P$-like levels albeit at a
different (larger) value of $V_{sp}$ as it is summarized in
Table~\ref{tab:02}. It appears that the mechanism of the level
reversal does not depend on the termination, 
and moreover, with a larger size of QD,
the splittings, $\Delta_S$ and $\Delta_P$ decrease.

We now turn our attention to the passivation of the QD surface. If we
consider the unpassivated 14-layer $\alpha$-type Ge QD, we find that
it is difficult to identify the band gap in its spectrum. There is a
gap in the sequence of energy levels as shown in Fig.~\ref{fig:C1}(b)
but it is shifted with respect to the bulk band gap (shown by dashed
lines) and more importantly, the wavefunctions of the states at the
anticipated top of the VB (and bottom of the CB) are localized at the
QD surface rather than extending through the volume of the QD. Since
surface passivation is known to be important both from the perspective
of enhancing measured optical 
properties\cite{Salivati2009:JAP,Yang2008:RLMS} and calculated positions
of energies in VB and CB,\cite{Hapala2013:PRB,Yu2006:APL} 
we have added an extra atomic layer
to the clusters which we investigated, making sure that there remains
no dangling bond of the QD core. The passivation layer atoms are
assumed to have the same hopping to the QD core atoms as the core
atoms between each other and to keep the model simple, we assign a
single on-site energy $E_{\rm pass}$ to their $s$- and $p$-orbitals. In our
convention, $E_{\rm pass}$ is taken relative to the on-site energy of QD
core atom $p$-orbitals. By
adjusting $E_{sp}$ sufficiently far from the bulk band gap region, we
remove the states localized at the surface from the gap and
recover the band gap in the QD spectrum as it can be seen in
Fig.~\ref{fig:C1}(c). For all calculations with passivated Ge QDs in
this article, we use $E_{\rm pass}=-5.41$~eV.

\setcounter{equation}{0}
\renewcommand{\theequation}{B-\arabic{equation}} 
\section*{Appendix D: Geometrical Factor in Energy Ordering}
\label{Apdx:geofactor}

\begin{figure}[htbp]
\includegraphics[scale=0.5]{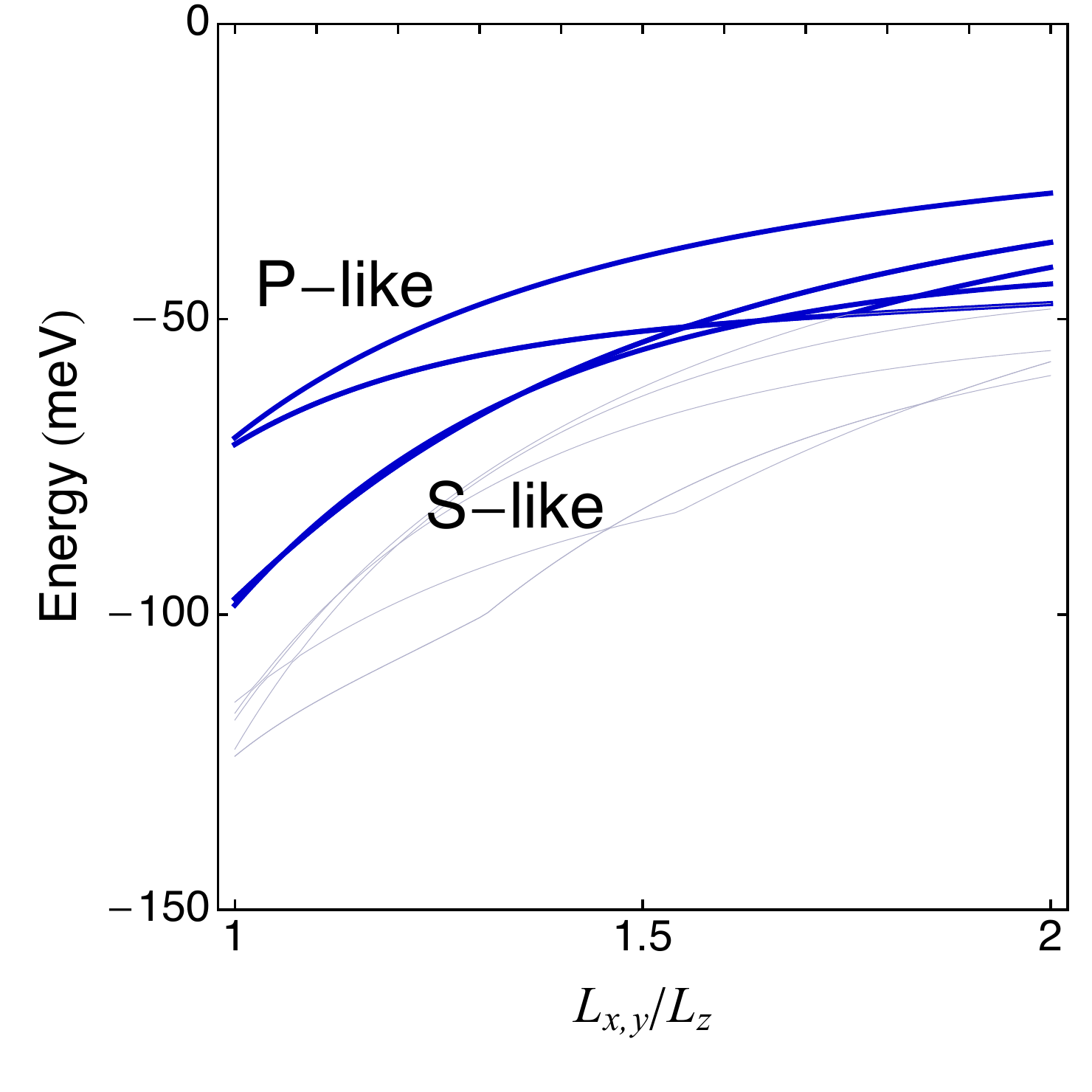}
\caption{
The evolution of  InAs QD VB energies as function of the aspect ratio. 
The lowest $P$-like state levels split %
for the aspect ratio different %
from unity. 
$L_z$ is fixed, while $L_x=L_y \in  [L_z$,$2L_z]$.
}\label{fig:D1}
\end{figure}

The geometry of actual QDs investigated in experiments 
deviates from a perfectly spheric or cubic shape. Our analysis has so far 
focused on a simple cubic geometry. Although the termination 
and passivation issues discussed in Appendix C already reduce 
the level of perfection of the QD shape, it is useful to estimate 
how geometrical changes in QD influence energy level ordering.

Using the same approach as in Ref. 61, we calculate levels in a
$L=9.8$ nm InAs QD of cuboid
shape $L_x=L_y\not= L_z$. Material parameters $\gamma_{1,2,3}$ for
this calculation were taken from Table~I (without spherical approximation) 
but $\Delta_{\rm{SO}}$ was set to zero. As
Fig.~\ref{fig:D1} shows, the originally threefold degenerate $S$-like and $P$-like
levels split as the aspect ratio $L_{x,y}/L_z$ deviates from one. The
top level of the VB is a $P$-like doublet and even though its separation
in energy from the lower-lying $S$-like state decreases with increasing
$L_{x,y}/L_z$, the two levels do not cross up to aspect ratios as
large as five. These results support the view of the $P$-like ground
state as a robust feature of QDs fabricated from a suitably chosen
material.

\setcounter{equation}{0}
\renewcommand{\theequation}{E-\arabic{equation}} 
\section*{Appendix E: Perturbative Analysis of the Tight-Binding Models}

\begin{figure}[htbp]
\includegraphics[scale=0.35]{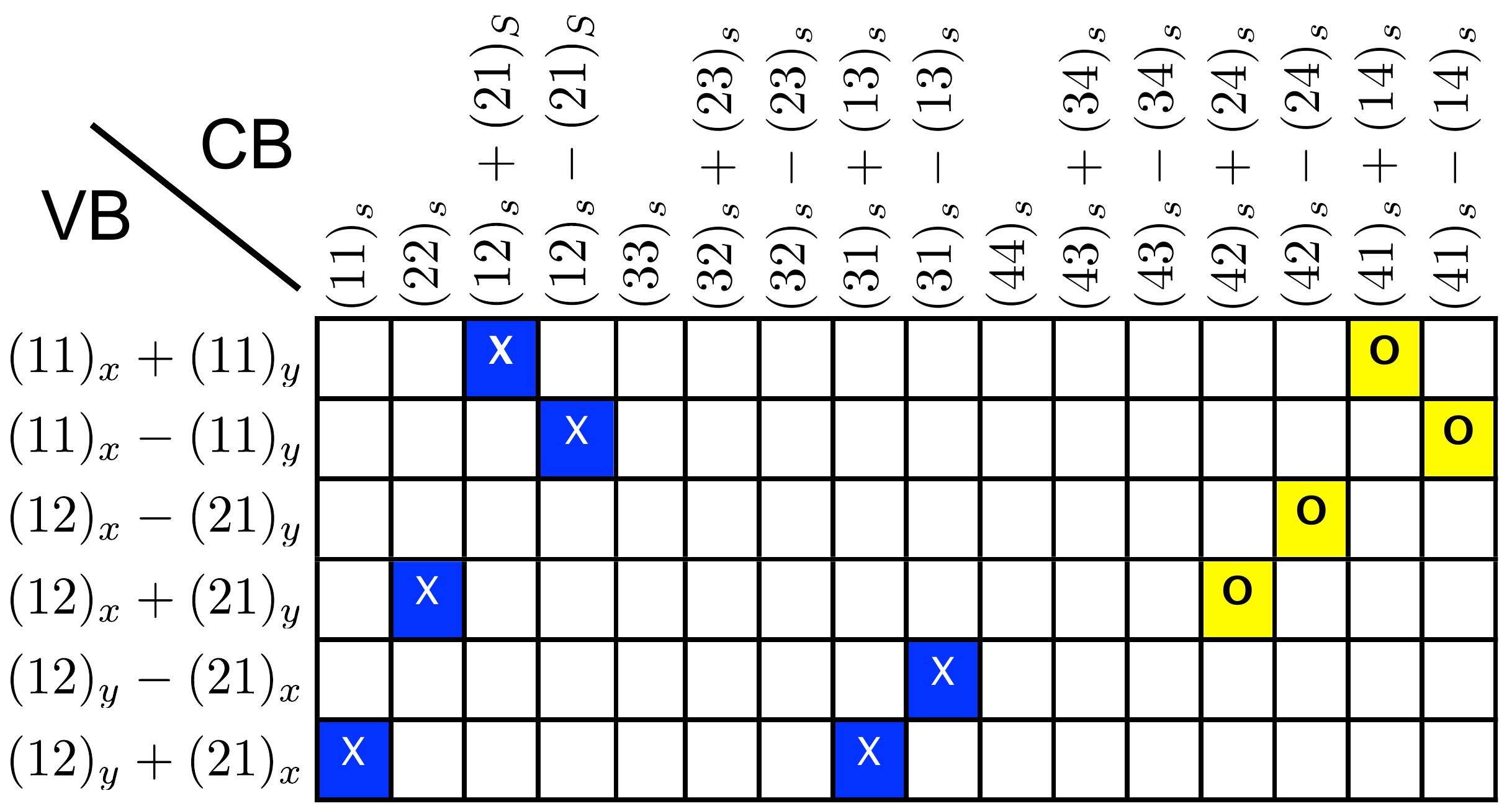}
\caption{Couplings of the six top VB states to CB states. The first
two rows correspond to $S$-like states, the other four to the
$P$-like states in 2D. Crosses (circles) indicate strong (weak) coupling, 
while blank cells corresponds to zero coupling.}\label{fig-01}
\end{figure}

We have already given an example of the coupling rules for $\Delta\hat{V}$
of Eq.~(\ref{eq:DeltaV})
between the VB and CB states in Eq.~(\ref{eq-03}). For a general CB
state $(n_xn_y)_s$ and a general VB state $(n_xn_y)_x$ or
$(n_xn_y)_y$, the matrix elements of $\hat{H}_{sp}$ defined in Eq.~(\ref{eq:Hsp}) can 
be written down analytically. They take on a particularly simple form in 
the continuum ($N\to\infty$) limit,
\begin{equation}\label{eq:E1}
  \begin{split}
\langle (nm)_x|\hat{H}_{sp}|(n'm')_s\rangle =& 8V_{sp}\delta_{mm'} A_{nn'}, \\
\langle (nm)_y|\hat{H}_{sp}|(n'm')_s\rangle =& 8V_{sp}\delta_{nn'} A_{mm'},
  \end{split}
\end{equation}
where $A_{nn'}=0$ if $n$ and $n'$ are both odd or both even and 
$nn'/((n')^2-n^2)$ otherwise.

We focus on the two $S$-like states given in Eq.~(\ref{eq:13tmp}) and four 
$P$-like states given in Eq.~(\ref{neq:13})
that are degenerate when $V_{sp}=0$. After a transformation applied to
to the $S$-like states, the six VB states of our interest can be listed as 
shown in the left column of Fig.~\ref{fig-01}. CB states can also be represented
in an analogous manner in order to make our analysis clear. 
It turns out that most matrix elements of $\Delta\hat{V}$ between
these six VB states and CB states vanish. 
In Fig.~\ref{fig-01},
large (small) non-zero matrix elements for the given pair of VB and CB
states are indicated by a cross (circle). If we choose a VB state, we can 
determine how strongly its energy depends on $V_{sp}$ by inspection 
of the corresponding row in the table. To the second order of the
perturbative analysis, the energy correction is proportional to $V_{sp}^2$
multiplied by the sum of squared matrix elements. We can convince ourselves
that in five of the six rows of Fig.~\ref{fig-01}, there is always at least one
large matrix element. The row which contains only small matrix
elements corresponds to the state $(12)_x-(21)_y$ which is
the OVS of Eq.~(\ref{eq-02}). The consequent weak $V_{sp}$-dependence of this state's energy is at the root of the level crossing in Fig.~\ref{fig:07}(d).

The weak interaction of OVS with the CB levels
is due to OVS symmetry which renders the couplings to the
low-$n_x,n_y$ CB states zero. The lowest CB state to which OVS
couples is $(42)_s-(24)_s$ and because of the large $n_x$ and $n_y$
involved in this state, the coupling matrix element in
Eq.~(\ref{eq:E1}) is relatively small compared to the couplings in
other rows of Fig.~\ref{fig-01}.


\begin{thebibliography}{10}
\bibitem{note2} Our basis in which we write down the matrix of the
Hamiltonian in Eq.~(\ref{eq:lutt}) comprises direct products of the $J_z$
eigenstates and envelope functions 
$\sin(n_x\pi x/L_0)\sin(n_y\pi y/L_0)\sin(n_z\pi z/L_0)$.
The unperturbed $S$-like wavefunction corresponds to $n_x=n_y=n_z=1$.
Throughout the paper $L_0=L$, except for Fig.~\ref{fig:03}(b) with 
well-in-well structure, where a smaller QD ($L\times L\times L$) structure is
confined by finite potential $V_0$ in a larger infinite potential well ($L_0\times L_0\times L_0$)
with $L_0=2L$.
\bibitem{note1} Note that the $S$-like manifold is precisely
$\{(11)_x,(11)_y\}$ only at $V_{sp}=0$. At finite $V_{sp}$, the two
wavefunctions acquire admixtures from other states defined by
Eq.~(\ref{eq:03}) but they remain degenerate. This degeneracy is due
to the $90^\circ$-rotation symmetry.
\end{thebibliography}
\end{document}